\documentclass[usenatbib,usedcolumn]{mnras/mnras}
\usepackage[utf8]{inputenc}
\usepackage[T1]{fontenc}
\usepackage{graphicx}
\usepackage[dvipsnames]{xcolor}
\usepackage[flushleft]{threeparttable}

\newcommand{\msun} {\(\textup{M}_\odot\) }
\newcommand{\mmsun}{\textup{M}_\odot}

\begin{document}

\title[Hierarchical fragmentation in high redshift galaxies]{Hierarchical fragmentation in high redshift galaxies revealed by hydrodynamical simulations}

\author[B. Faure et al.]{
Baptiste Faure$^{1}$\thanks{Contact e-mail: \href{mailto:baptiste.faure@cea.fr}{baptiste.faure@cea.fr}},
Frédéric Bournaud$^{1}$,
Jérémy Fensch$^{2,3}$,
Emanuele Daddi$^{1}$,
\newauthor{
Manuel Behrendt$^{4,5}$,
Andreas Burkert$^{4,5}$
and Johan Richard$^{6}$}
\\
$^{1}$AIM, CEA, CNRS, Université Paris-Saclay, Université Paris Diderot, Sorbonne Paris Cité, 91191 Gif-sur-Yvette, France\\
$^{2}$Univ. Lyon, ENS de Lyon, Univ. Lyon 1, CNRS, Centre de recherche Astrophysique de Lyon, UMR5574, F-69007 Lyon, France\\
$^{3}$European Southern Observatory, Karl-Schwarzschild-Str. 2, 85748 Garching, Germany\\
$^{4}$Universitäts-Sternwarte München, Scheinerstr. 1, D-81679 München, Germany\\
$^{5}$Max Planck Institute for Extraterrestrial Physics, Giessenbachstraße 1, D-85748 Garching, Germany\\
$^{6}$Univ Lyon, Univ Lyon1, Ens de Lyon, CNRS, Centre de Recherche Astrophysique de Lyon UMR5574, F-69230, Saint-Genis-Laval, France
}

\maketitle

\begin{abstract}

High-redshift star-forming galaxies have very different morphologies compared to nearby ones. Indeed, they are often dominated by bright star-forming structures of masses up to $10^{8-9}$ \msun dubbed «giant clumps». However, recent observations questioned this result by showing only low-mass structures or no structure at all.
We use Adaptative Mesh Refinement hydrodynamical simulations of galaxies with parsec-scale resolution to study the formation of structures inside clumpy high-redshift galaxies. We show that in very gas-rich galaxies star formation occurs in small gas clusters with masses below $10^{7-8}$ \msun that are themselves located inside giant complexes with masses up to $10^8$ and sometimes $10^9$ \msun. Those massive structures are similar in mass and size to the giant clumps observed in imaging surveys, in particular with the Hubble Space Telescope.
Using mock observations of simulated galaxies, we show that at very high resolution with instruments like the Atacama Large Millimeter Array or through gravitational lensing, only low-mass structures are likely to be detected, and their gathering into giant complexes might be missed. This leads to the non-detection of the giant clumps and therefore introduces a bias in the detection of these structures. We show that the simulated giant clumps can be gravitationally bound even when undetected in mocks representative for ALMA observations and HST observations of lensed galaxies. We then compare the top-down fragmentation of an initially warm disc and the bottom-up fragmentation of an initially cold disc to show that the process of formation of the clumps does not impact their physical properties.
\end{abstract}

\begin{keywords}
galaxy evolution -- high redshift
\end{keywords}

\section{Introduction}

For two decades, deep imaging surveys have revealed that high-redshift ($z$>1) star-forming galaxies have optical and near-infrared morphologies that strongly differ from nearby galaxies. The distribution of star formation that is revealed by optical data is often dominated by irregular structures such as the so-called ``giant clumps’’ and at the same time long spiral arms are often absent \citep[e.g.,][]{Cowie96, Elmegreen05, Genzel06, Guo18, Zanella19}. The giant clumps can reach sizes of several hundreds of parsecs and stellar masses of a few $10^8$, sometimes $10^9$, solar masses. The clumps are actively star-forming with typical star formation rates of several solar masses per year in each clump \citep{EE05, Elmegreen07}. Clumpy galaxies are generally found to have mass distributions and velocity field consistent with rotating discs \citep{Genzel08, Bournaud08} and generally lack signatures of mergers \citep{Cibinel15}. The standard picture for the formation of such clumpy disc galaxies is the fragmentation, under gravitational instability, of gas-rich discs \citep{Noguchi99, BEE07, Agertz09, DSC09, Ceverino10}. The high observed gas fractions, about 50\% of the baryonic mass at redshift 2 \citep{Daddi10,Tacconi10, Combes13,Santini14,Zanella18} and high turbulent speeds \citep[$\sim$50\,km\,s$^{-1}$][]{Genzel06, Genzel08, Bournaud08, Swinbank09, Swinbank10} are consistent with this scenario. In this scenario, giant clumps form by gravitational instabilities and may subsequently fragment in sub-structures while remaining gravitationally bound. Alternatively, a \textit{bottom-up} scenario where the gaseous disc fragments into small clumps that then agglomerate to form the giant clumps is also proposed \citep{Behrendt15, Behrendt16, Behrendt19}. To date, observations lack resolution to detect the predicted substructures and thus understand their formation.

The evolution of giant clumps remains actively debated. In particular no consensus has been found on their survival against stellar feedback. Some recent simulations \citep[e.g., ][but see \citealt{Fensch20}]{Hopkins12,Oklopcic17} suggest that feedback is strong enough to destroy the giant clumps in a few tens of million years. Indeed radiative feedback from massive stars may play an important role in the disruption of giant clumps \citep{Murray10, Krumholz10}. Yet, other analytical models and simulations including radiative feedback found that feedback effects can disrupt clumps less massive than $10^{7-8}$\,M$_{\sun}$, while more massive clumps could survive feedback, due in particular to the continuous re-accretion of gas from the large-scale reservoirs in the galactic disc \citep{Dekel13, Bournaud14, Ceverino15}. Note that survival to feedback makes it possible for giant clumps to migrate toward the center of the galaxy through dynamical friction, leading to the growth of the bulge \citep[e.g.,][for a review]{Bournaud16}. Probing the evolution of giant clumps remains crucial to understand the evolution of high-redshift galaxies.\\

More recently, the very existence of giant clumps themselves has been questioned. Indeed, observations of strongly lensed galaxies, such as the Cosmic Snake \citep{Cava18}, only detected smaller and lower mass clumps, with a median stellar mass not larger than $1\times10^8$ \msun yielding to the conclusion that the mass of giant clumps in Hubble Space Telescope (HST) images could have been largely overestimated \citep[see also][]{Dessauges17, Dessauges19}. Moreover attempts to detect the gaseous counter part of optical giant clumps with Atacama Large Millimeter Array (ALMA) \citep{Cibinel17, Rujopakarn19, Ivison20} only provided upper limits on gas masses lower than those expected from the UV rest frame luminosity and star formation rate of the giant clumps. This might indicate that gas has already been expelled by feedback from stellar giant clumps that are still UV bright.
These observations also tentatively question the survival to feedback if not the very existence of giant clumps in the $10^8$ and $10^9$ \msun mass range and sizes up to $\sim 500$ pc by proposing, inter alia, beam smearing or differential extinction as explanation of previously mis-interpreted observations. \\

In this paper we show that the observations at high \textit{effective} resolution (ie. high angular resolution or observation of strongly lensed galaxies) detect the sub-structures of the giant clumps but may miss these giant clumps. We use sub-parsec scale numerical simulations of isolated galaxies, focusing on purpose on very gas-rich and very clumpy systems, as described in Section 2. From those, we create mock observations whose wavelength and resolution are representative of previously cited observations as well as integration of gravitational lens model. We specify the mocks creation in Section 3. We then analyse the clumps found on the mocks by a clump-finder in Section 4. We then discuss the connection between all scales structures before analysing their physical existence as well as their process of formation in Section 5 and 6. Finally, we show that detecting the giant clumps in high resolution data would require to degrade the resolution, which is not possible with existing ALMA data that generally lack sensitivity.

\section{Simulations}

Our goal is to simulate isolated  gas-rich ($\sim 50\% $ gas fraction) galaxies with little bulge ($\sim 15\%$), which is not uncommon at a redshift of 2. There exist a variety of galaxies at those redshift but we focus here on the most disk-dominated and gas-rich ones, and the most clumpy ones. Less clumpy galaxies could result from the same physical model -- in particular a similar feedback scheme -- with a somewhat lower gas fraction (see below and \citealt{Fensch20}) or, possibly, a higher bulge/spheroid fraction \citep{BE09}.

\begin{table*}
\caption{Simulations parameter}
\begin{center}
\begin{tabular}{lcccc}
\hline
Parameter & Simulation 1 & Simulation 2 & Simulation 3 & Simulation 4 (Behrendt et al., in prep) \\ \hline \hline
Initial gas mass & $2.0 \times 10^{10} \mmsun $ & $3.5 \times 10^{10} \mmsun $ & $6.4 \times 10^{10} \mmsun $ & $3.7 \times 10^{10} \mmsun $ \\ 
Initial gas fraction & 50\% & 50\% & 50\% & 100\%\\
Gaseous exponential radius & 5 kpc & 13 kpc & 12 kpc & 5.26 kpc \\ 
Stellar exponential radius & 5 kpc & 4 kpc & 5 kpc & 5.26 kpc \\ 
Feedback model & Full & Full & Kinetic only & SN feedback only\\ 
Cooling model & Full & Full & Pseudo & Full with temperature floor at $10^4$K \\ 
Fine AMR resolution & 1.5pc/cell & 0.4pc/cell & 0.2pc/cell & 2.9pc/cell \\ 
Coarse AMR resolution & 780pc/cell & 390pc/cell & 195pc/cell & 750pc/cell \\ 
Star formation threshold & $1.0\times10^2$ H/cc & $1.0\times10^2$ H/cc & $1.0\times10^4$ H/cc & $2.0\times10^4$ H/cc\\ 
Mass of new stars & $1.2\times10^4$ \msun & $1.5\times10^3$ \msun & $1.9\times10^4$ \msun & $2.3\times10^4$ \msun\\
Bulge mass in stellar mass fraction & 15\% & 15\% & 15\% & 0\% \\ 
\hline
\end{tabular}
\end{center}
\label{table:simu_param}
\end{table*}%

\subsection{Simulation technique}
\label{sec:simutechnique}

The simulations presented in this paper are performed with the Adaptative Mesh Refinement (AMR) code RAMSES \citep{Teyssier02} with physical models globally similar to in \cite{Bournaud14}. The coarse level ranges from 200 to 800~pc for each simulation, and each AMR cell is refined into $2^3$ new cells if i) its gas mass is larger than $2\times10^4 \: \mmsun $, or ii) the local thermal Jeans length is smaller than four cells, or iii) it contains more than 40 particles. The smallest resolution ranges from 1.0 to 0.2~pc (see Table \ref{table:simu_param} for each simulation). An artificial pressure floor is added to high-density gas, such that the Jeans' length cannot drop below four time the smallest cell size. This is typically considered to avoid artificial fragmentation by accounting for stabilising internal turbulent motions, smeared out by the resolution \citep{Truelove97,Ceverino12}. As in \cite{Teyssier10}, the equation of this pressure floor reads, with $\mathrm{x_{min}}$ being the smallest cell size :
\begin{equation}
    P_{Jeans} = 16 \mathrm{x_{min}^2}G\rho_{gas}^2/\gamma\pi
\end{equation}

One should note that this artificial pressure floor does not impact fragmentation and clumps properties. Indeed, turbulence dominates over thermal pressure in the studied clumps (see Section \ref{sec:structuresformation}). As an additional precaution we impose a minimal size for our clumps, in order not to have structures that are only on the pressure floor.

The simulations start as idealised models of isolated galaxies with sizes, masses and gas fractions representative of star-forming galaxies at redshift $z \sim 2$ . Table \ref{table:simu_param} lists the initial parameters of our three galaxy models. A fourth model from Behrendt et al. (in prep) is also used and is labelled as Simulation 4. It was also performed with RAMSES and presents the same artificial pressure floor as in our simulations.

The stellar sizes and masses are representative for the typical mass-size relation of high-redshift galaxies \citep{Dutton11}. Over the simulations we vary the gas disc scale-length with respect to the stellar disc radius to have a proper sample of galactic compacity\footnote{While the gaseous scale length are largely unknown at redshift 2, such a diversity of gas compactness with respect to stellar sizes is observed at least in the local universe \citep{deBlok08}.}. Simulations 1 to 3 all start with a stellar bulge that represents 15\% of the initial stellar mass.

Simulations 1 to 3 start with an initial disc at a temperature of $2\times10^5$ K, preventing it to form structures. The disc evolves for 100Myr at this constant and warm temperature so that the initial conditions relax into an axisymmetric disc at equilibrium. After this initial phase, gas cooling is activated, which allows gas to form dense structures. In simulations 1 and 2, the cooling is done by successive stages to assure the first structures to form are massive. At the time of the analysis, our three original simulations were run for a few hundred million years, the time for the feedback activation plus a few dynamical times. Simulation 4 was run for 700 Myr, the time for the galaxy to reach a 73\% gas fraction.

Cooling is implemented similarly to \cite{Perret14}: we model fine-structure cooling, heating from a \citet{Haardt96} uniform UV background, the cooling and heating rates being tabulated by \cite{Courty04} assuming solar gas  metallicity.

\subsection{Star formation and feedback}
Star formation and feedback are modelled as in \cite{Renaud13} \citep[see also][]{Dubois08}. At each time step of duration d$t$, cells with gas density higher than a threshold $\rho_*$ are allowed to form stars. For each cell above the threshold and of physical size $d_x$, a dimensionless integer $n_*$ is drawn, following a Poisson distribution of mean value $\rho_{\mathrm{SFR}}d_x^3\mathrm{d}t/M_*$: a non-zero value of $n_*$ implies the conversion of the mass of $n_*M_*$ into one or a few stellar particles. $\rho_{\mathrm{SFR}}$ is the local star formation rate according to the Schmidt law: $\rho_{\mathrm{SFR}} = \epsilon\rho/t_{\mathrm{ff}}$, where $\rho$ is the gas density, $\epsilon$ is the star formation efficiency set to 2\% throughout this paper, and $t_{\mathrm{ff}}$ is the local free-fall time given by $t_{\mathrm{ff}} = \sqrt{3\pi/(32G\rho)}$. $M_*$ is the mass of newly formed stellar particles.

Three different stellar feedback mechanisms have been included, following \cite{Renaud13} and \cite{Bournaud14}:
\begin{itemize}
\item \textit{photo-ionization of HII regions}: around each stellar particle younger than 10~Myr, a photoionized region is computed using a Strömgren sphere approximation, taking into account recombination. Each sphere can be larger or smaller than a gas cell. When HII regions overlap, the volume of each is increased not to ionize the gas twice. 
Gas in this HII region is heated to $2.5\times10^4$ K to model photo-ionization. We take into account the doubling of number density through photo-inoization.More details can be found in \cite{Renaud13}.

\item \textit{Radiation Pressure}, using the scheme described in \cite{Renaud13}: a fraction of the momentum available in photons emitted by young stars is distributed to the gas in the HII regions defined above. The momentum is time-dependant and computed from the luminosity of stars younger than 10 Myr (see equation 2 in \cite{Renaud13}.

\item \textit{Supernovae}: as in \cite{Bournaud14}, 20\% of the mass of stellar particles is converted into energy into the surrounding gas 10 Myr after their formation in the form of kinetic (20\%) and thermal energy (80\%). The kinetic energy is injected in a 3 cell-radius sphere, in the form of a velocity kick.
\end{itemize}

It is important to note that these feedback recipes and the calibration used here are not particularly weak (and the cooling is not particularly strong), in the sense that, when used in gas-poor galaxies they are not creating clumps as large or massive as the one that will be studied here and the lower-mass clump formed in gas poor galaxies are not long-lived. The same set of feedback recipes and a roughly similar calibration has been used in \citep{Renaud15} and successfully created short-lived Giant Molecular Clouds in isolated (and interaction) Milky Way-like galaxies with 5--10\% gas fractions.
Without going to such low gas fractions (that are rare are redshift two among star-forming galaxies), \citep{Fensch20} have recently shown that lowering the gas fraction to 25\%, that is by only a factor two (which is not uncommon among star-forming galaxies at redshift two) results in clumps that are slightly less massive and are above all much shorter-lived, independently of the very details of the feedback calibration. Hence out simulations are expected to represent a large fraction of star-forming disk-dominated galaxies at redshift two, but do not imply that all of them are extremely clumpy, even less at lower redshifts. Note that the formation of giant clumps with a gas fraction about 50\% also depends on the bulge/spheroid mass \citep{BE09}.

The first two simulations were run with this whole set of feedback mechanisms. The third simulation uses a simpler model in order to reach higher spatial resolution. Instead of computing gas cooling and heating we impose an equation of state where the temperature is defined as a function of gas density (the \textit{pseudo-cooling} equation of state, \cite{Bournaud10}, Figure 1) while keeping the pressure floor introduced in the previous section. Only kinetic supernovae and radiation pressure feedback are used in this third simulation.

In addition, we checked that the star formation rate of each galaxy is representative for main sequence galaxies at a redshift of two. The simulations properties at the time of analysis are summarised in Table \ref{table:simu_analysis}.
Each galaxy presents a SFR between 30 and 215 $\mmsun\:\textup{yr}^{-1}$, consistent with star forming galaxies of similar stellar masses at redshift 2 \citep{Elbaz2011,Schreiber15}.

\begin{table*}
\caption{Simulations parameter at the time of the analysis}
\begin{threeparttable}

\begin{center}
\begin{tabular}{lcccc}
\hline
Parameter & Simulation 1 & Simulation 2 & Simulation 3 & Simulation 4 \\ \hline \hline
Gas mass & $9 \times 10^{9} \: \mmsun $ & $3.1 \times 10^{10} \: \mmsun $ & $2.6 \times 10^{10} \: \mmsun $ & $2.3 \times 10^{10} \: \mmsun $ \\ 
Gas fraction & $23 \: \%$ & $44 \: \%$  & $20 \: \%$ & $73 \: \%$\\
Stellar mass & $3.1 \times 10^{10} \: \mmsun$ & $3.9 \times 10^{10} \: \mmsun$ & $1.0 \times 10^{11} \: \mmsun$ & $8.2 \times 10^{9} \: \mmsun$ \\
Star Formation Rate \tnote{*} & $75 \: \mmsun \cdot \mathrm{yr}^{-1}$ & $79 \: \mmsun \cdot \mathrm{yr}^{-1}$ & $215 \: \mmsun \cdot \mathrm{yr}^{-1}$ & $33 \: \mmsun \cdot \mathrm{yr}^{-1}$  \\
Specific Star Formation Rate & $2.4 \: \mathrm{Gyr}^{-1}$ & $2.0 \: \mathrm{Gyr}^{-1}$ & $2.2 \: \mathrm{Gyr}^{-1}$ & $4\: \mathrm{Gyr}^{-1}$\\
Time at analysis & 340 Myr & 346 Myr & 200 Myr & 740 Myr\\
\hline
\end{tabular}
\begin{tablenotes}
\item[*]SFR is computed by averaging the total gas diminution over 15Myr.
\end{tablenotes}
\end{center}

\end{threeparttable}

\label{table:simu_analysis}
\end{table*}%

\section{Analysis}
\subsection{Generating mock HST images}
\label{MockObsTh}

We create mock HST observations in the F814W filter assuming a redshift $z=2$. We use the \cite{Bruzual03} stellar evolution model, with solar metallicity and \cite{Salpeter55} initial mass function (IMF).\footnote{Mocks created with a \cite{Chabrier03} IMF only present a higher contrast between low and high stellar mass regions as seen in Appendix \ref{app:IMF}. The clumps detection being mainly determined by the clump finder parameters (see \ref{sec:clumpfinder}), the IMF does not play a significant role here.} Stars present in the initial conditions are given a random age between 300 Myr and 3 Gyr with a uniform distribution.

Luminosity maps are created at two different resolutions: one matching HST resolution in the band F814W for $z = 2$, i.e. a 0.05" angular resolution corresponding to roughly 500 parsec per pixels, and one with a very high resolution of 12.2 parsec per pixels. The first maps are created to compare with HST observations, and the latter in order to detect possible smaller-scale structures.

Dust attenuation is similar in giant clumps and in the rest of clumpy galaxies \citep{Elmegreen05, Elmegreen07}. We do not take into account dust attenuation or scattering for the creation of the mocks. However, previous studies performed with dust extinction present similar irregular and clumpy structures \citep[see in particular][]{Behrens18}.

\subsection{ALMA}
\label{sec:ALMAOST}

We use the ALMA Observation Support Tool \citep[OST,][]{Heywood11} to create mocks representative for ALMA dust continuum observations that include sensitivity effects. We compute an integrated infra-red luminosity from the star formation rate assuming a Salpeter IMF \citep{Salpeter55} using \cite[][Equation 8]{Bethermin12} model. From the infra-red luminosity we compute the flux for a galaxy at $z$\,=\,2 observed at 217\,GHz assuming the Spectral Energy Distribution from \cite{Bethermin12} with a dust temperature of 30K and in agreement with \cite{Magdis12, Bethermin15}. As a based map for ALMA OST, we use a map of the gas denser than 100\,H/cc.
We then use the OST simulator with different beam sizes and different observation times with a precipitable water vapor (PWV) of 0.472mm. As a reference, we also create an ideal observation with virtually infinite sensitivity. Those mocks are used in Section \ref{sec:ALMAthreshold} in order to tackle sensitivity effects.

\subsection{Creation of the lensed images}

We have produced realistic observations of the simulated maps as seen through the magnification of a lensing cluster core. To do this, we have used the software Lenstool \citep{Jullo07} \footnote{Publicly available at \url{https://projets.lam.fr/projects/lenstool/wiki}} and the model constructed for the cluster MACSJ1206 \citep{Ebeling09}, where the clumpy $z=1$ arc called the ‘Cosmic Snake’ \citep{Cava18, Dessauges19} is located. By applying high resolution \textit{displacement maps}, giving the angular deflexion at a given point in the image plane, to our simulation, and surface brightness conservation, we are able to reproduce image deflexion, magnification, as well as multiplicity. The instrumental PSF is afterwards applied on these simulated images. We have adjusted the source plane location of our simulated galaxy in order to produce a 22 arcsec long extended arc, similar to the Cosmic Snake. At this location, the typical magnification ranges from $\mu=4-28$, resulting in an effective resolution between $\sim 125$ pc and $\sim 20$ pc for HST mock observation. This galaxy is selected as it exhibits a direct comparison of a typical main-sequence high-redshift galaxy to its strongly lensed counterpart. Moreover, its lensing is stronger than most lensed galaxies allowing for a much better analysis of the structures.

\subsection{Clump finder}
\label{sec:clumpfinder}

In order to detect structures and clumps we use the clump finder Astrodendro \footnote{http://www.dendrograms.org/}. The algorithm creates a dendogram starting from the pixel with the maximum value. It then goes to the next largest value. If this pixel is a local maximum a new structure is created and if it is not, it is added to the closest structure. Two structures are merged into a branch as soon as the selected pixel is not a local maximum and is adjacent to two structures. One can take into account the noise of the data by setting a threshold for both the minimal value (min\_value) and the minimal interval required between two peaks to create a new structure (min\_delta). One can also select the number of pixels a structure needs to have to be considered as a clump (min\_npix). Each value of min\_npix and min\_value are selected so that found structures cannot be only composed of cells at the finest refinement level. This is done in order to avoid having strong artificial thermal pressure due to the high-density pressure floor (see Section \ref{sec:simutechnique}). In order to use the same parameters over all simulations, we defined them from the r.m.s. of the input image, depending on the resolution (see Table~\ref{table:dendro_param}).

\begin{table}
\caption{Clump finder parameter. Each value is multiplied by the rms of the input image.}
\begin{center}
\begin{tabular}{lccc}
\hline
Parameter & HST & Lensed HST & High resolution\\ \hline \hline
min\_value & 0.1 & 0.1 & 1000\\ 
min\_delta & 0.1 & 0.1 & 100\\ 
min\_npix & 3 & 5 & 20\\
\hline
\end{tabular}
\end{center}
\label{table:dendro_param}
\end{table}%

We run the clump finder on each type of mock images for all four simulations. Once a clump is found we compute the gas mass, the stellar mass and the mass of the stars that were formed less than $10^8$ years before the time of the mock, called young stars mass hereafter. 
The computation is done by a direct use of the simulation data to get the mass of each component contained in the contour of the clumps. As we do not detect three-dimensional structures the mass is integrated on the thickness of the disc. Using Astrodendro for a 3D detection of the structures shows only a 10\% difference in the masses values compared to the method described above. For simplicity and consistence with observations, the 2D detection is used in the next sections.

\section{Mock images and clump detection}

\subsection{HST clumps}

We define as "clumps" the structures found by Astrodendro and whose most intense pixel is located more than 2 kpc from the center of the galaxy, which is defined as the center of our simulation volume. This allows us to remove the central bulge and its structure from the analysis.

\begin{figure*}
\begin{center}
\includegraphics[width=\textwidth]{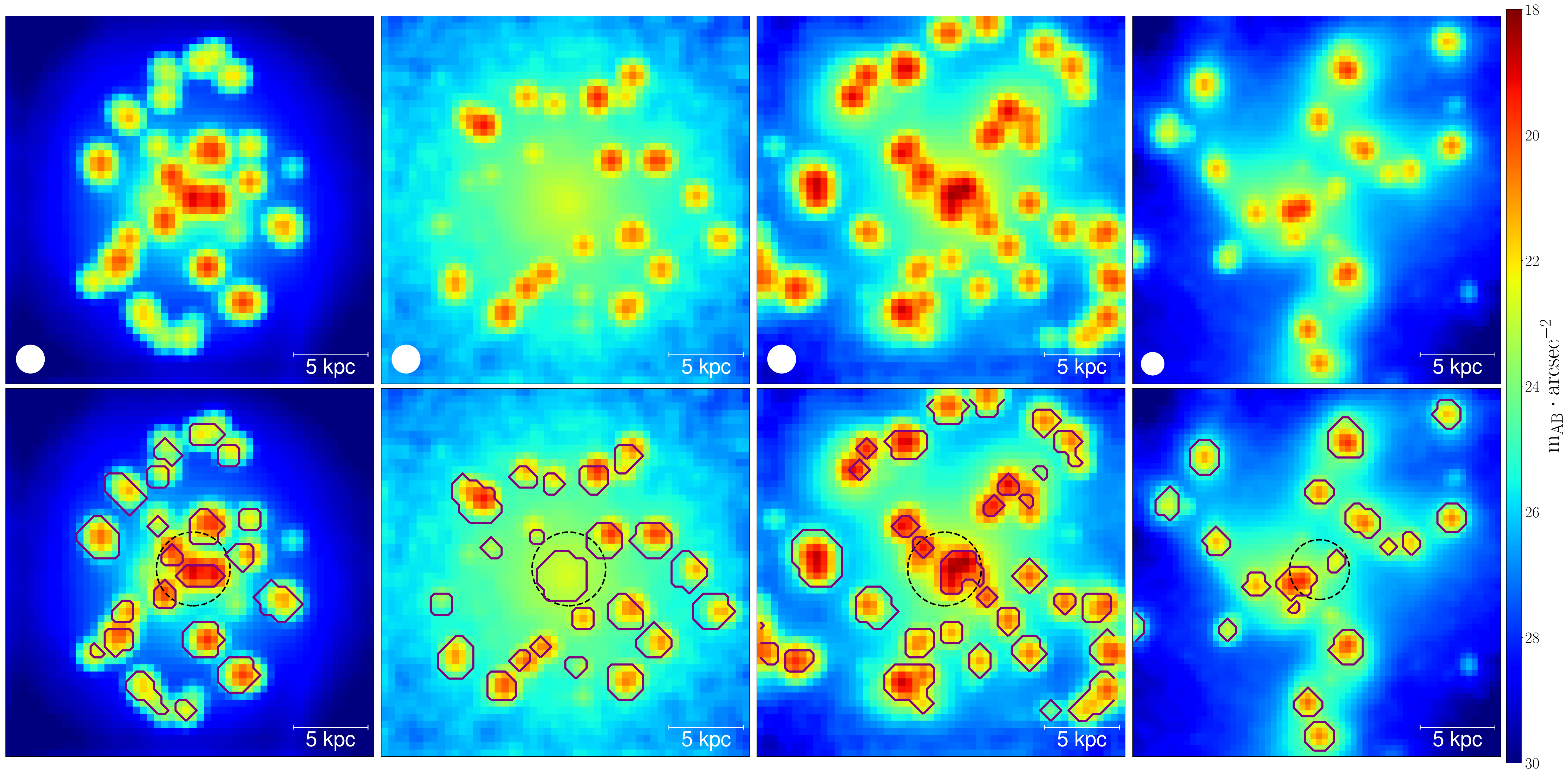}
\caption{Mock HST F814W observations of the simulated galaxies as if they were at z = 2. From left to right are the simulations 1 to 4. White patches on the bottom left corners are the size of the PSF.
The second row shows in red contours the clumps found by Astrodendro. The black dotted circles are the central 2 kiloparsec.
The mocks are created without any noise  nor dust attenuation therefore they can present clumps that will not be detected by the HST.}
\label{mockHST}
\end{center} 
\end{figure*}

\begin{figure}
\begin{center}
\includegraphics[width=\columnwidth]{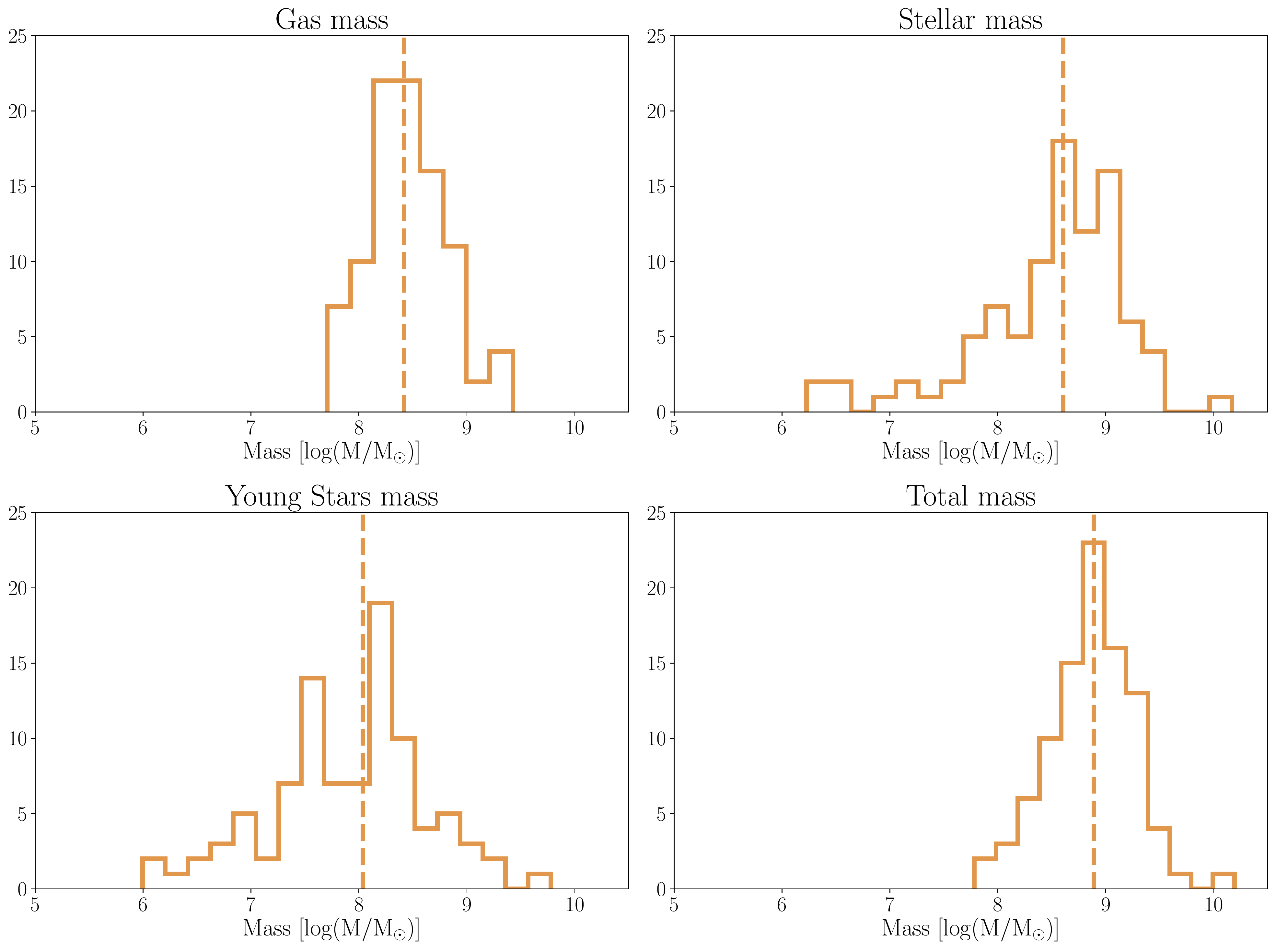}
\caption{Mass histogram of the HST-like clumps. A detail comparison simulation per simulation is shown in Figure \ref{bigplot}.}
\label{histHST}
\end{center}
\end{figure}

Figure \ref{mockHST} shows typical HST images obtained and the detected structures. One can see that every image contains a handful of structures with diameter around the kiloparsec, similarly to observations (see introduction). In Figure \ref{histHST} is shown the mass histogram of all the clumps found in all four simulations. The median gas mass is $2.6\times10^8$ \msun. The median stellar mass is $4.0\times10^8$ \msun. Roughly 25\% of the stellar mass in made-up of stars younger than 100\,Myr. The median total mass is $7.8\times10^8$ \msun. Those results are consistent with observations of giant clumps in the UV rest frame of high-redshift galaxies \citep[for example]{Guo18, Zanella19}.

As the parameters are computed directly from the simulation data and not from the emission, those results are not affected by the lack of dust attenuation: these are the intrinsic values of the structures.

For more detailed mass distributions, see Appendix \ref{bigplot}.

\subsection{HST lensed clumps}

\begin{figure*}
\begin{center}
\includegraphics[width=\textwidth]{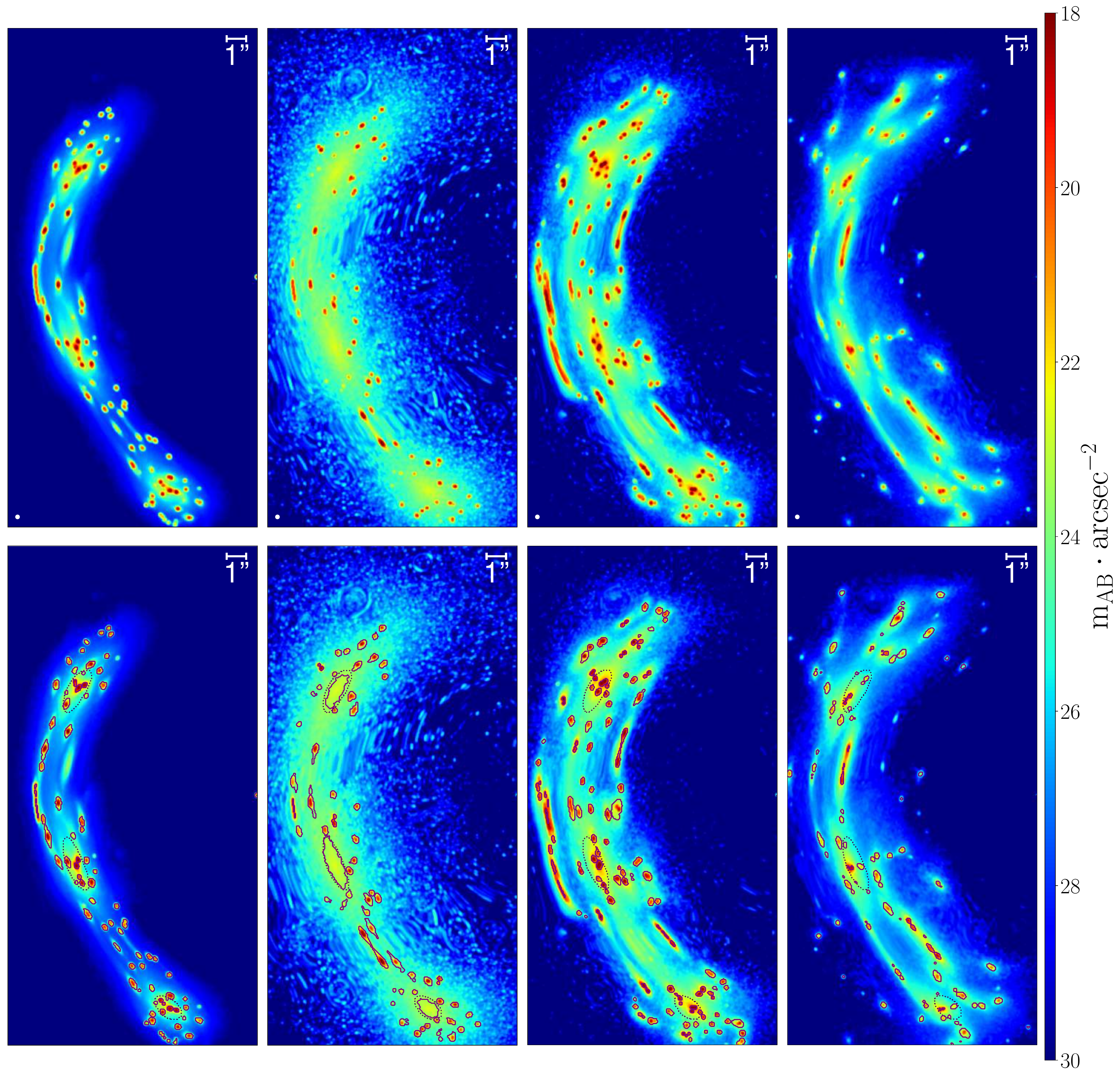}
\caption{Mock HST F814W observations of the simulated galaxies after the application of the lens model. The sources are at z = 2. From left to right are the simulations 1 to 4. White patches on the bottom left corners are the size of the PSF.
The second row shows in red contours the clumps found by Astrodendro.
The mocks are created without any noise nor dust attenuation therefore they can present clumps that will not be detected by the HST.}
\label{mockLens}
\end{center}
\end{figure*}

\begin{figure}
\begin{center}
\includegraphics[width=\columnwidth]{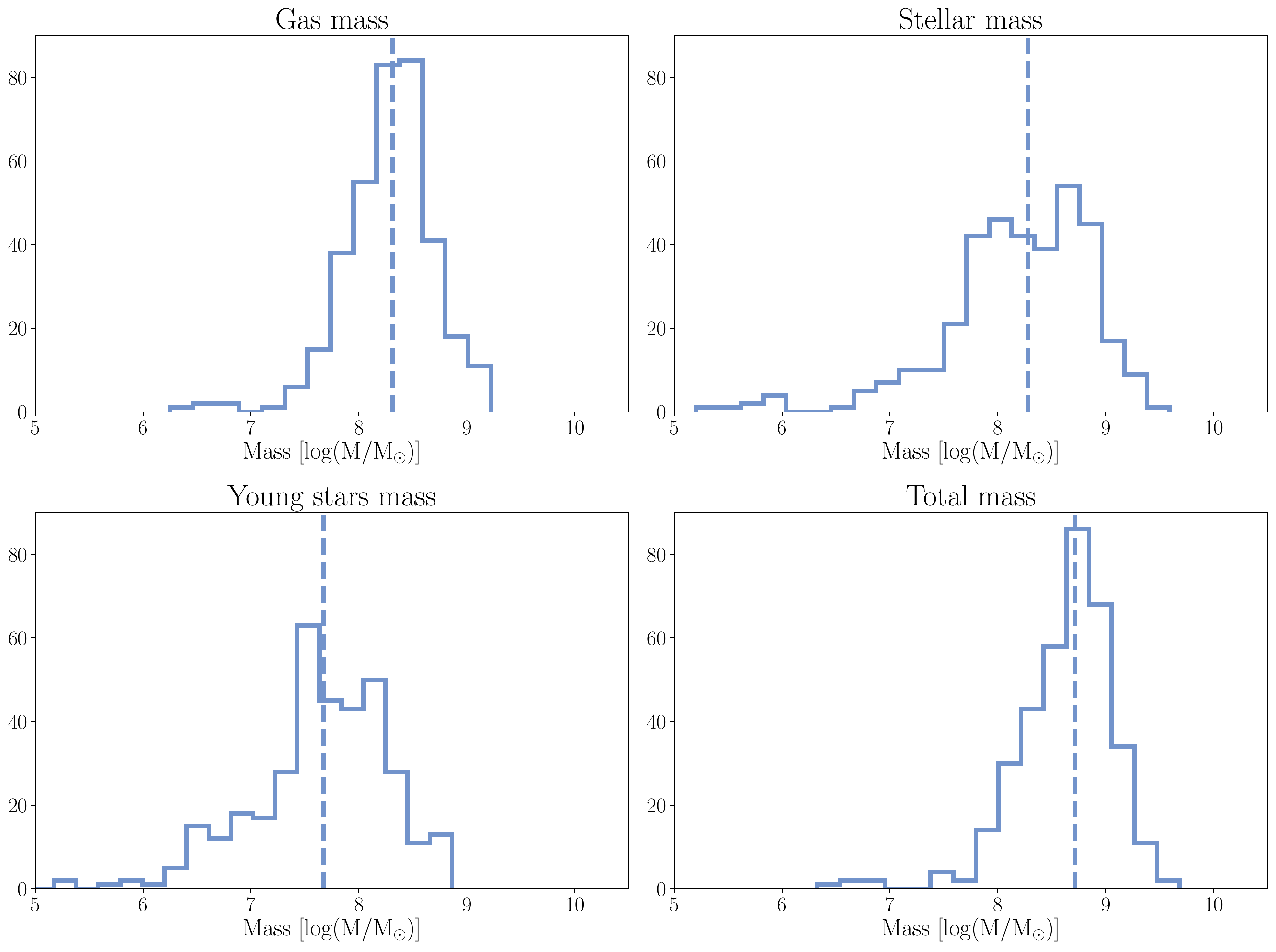}
\caption{Mass histogram of the HST-lensed-like clumps. A detail comparison simulation per simulation is shown in Figure \ref{bigplot}.}
\label{histLens}
\end{center}
\end{figure}

Figure \ref{mockLens} shows typical HST like lensed images obtained for the three different simulations and the detected structures by Astrodendro. One can see that every image contains much more structures than in the HST like non lensed images. The number and size are in agreement with observational studies such as \cite{Cava18}. Similarly to the previous section, Figure \ref{histLens} shows the mass histogram of all the clumps found in all four simulations. The median gas mass is $2.0\times10^8$ \msun. The median stellar mass is $1.9\times10^8$ \msun. Roughly 25\% of the stellar mass is made-up of stars younger than 100\,Myr. The median total mass is $5.2\times10^8$ \msun. Those masses are consistent with studies such as \citet{Dessauges17, Dessauges19} and \citet{Cava18}, where the observed galaxies present similar star formation rate and masses than our simulations.
The distribution of masses is reaching values as high as in the previous mocks as the resolution is not increased uniformly by the lensing, leading to a broad distribution. For  detailed mass distributions, see Appendix~\ref{bigplot}.

\subsection{High resolution clumps}

\begin{figure*}
\begin{center}
\includegraphics[width=\textwidth]{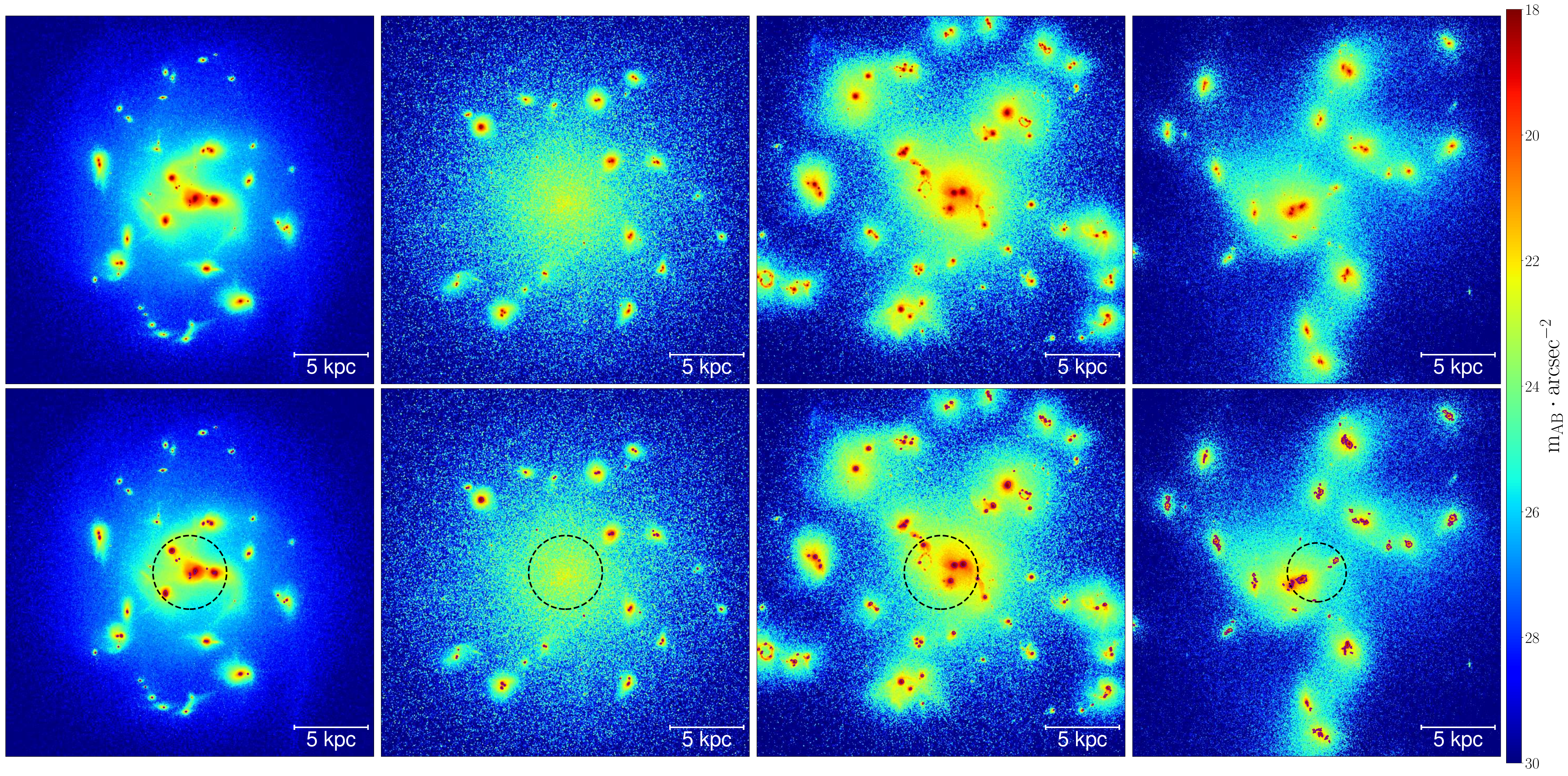}
\caption{High resolution images of the F814W image of the simulated galaxies. From left to right are the simulations 1 to 4. White patches on the bottom left corners are the size of the PSF.
The second row shows in purple contours the clumps found by Astrodendro. The black dotted circles are the central 2 kiloparsec.}
\label{mockHR}
\end{center}
\end{figure*}

\begin{figure}
\begin{center}
\includegraphics[width=\columnwidth]{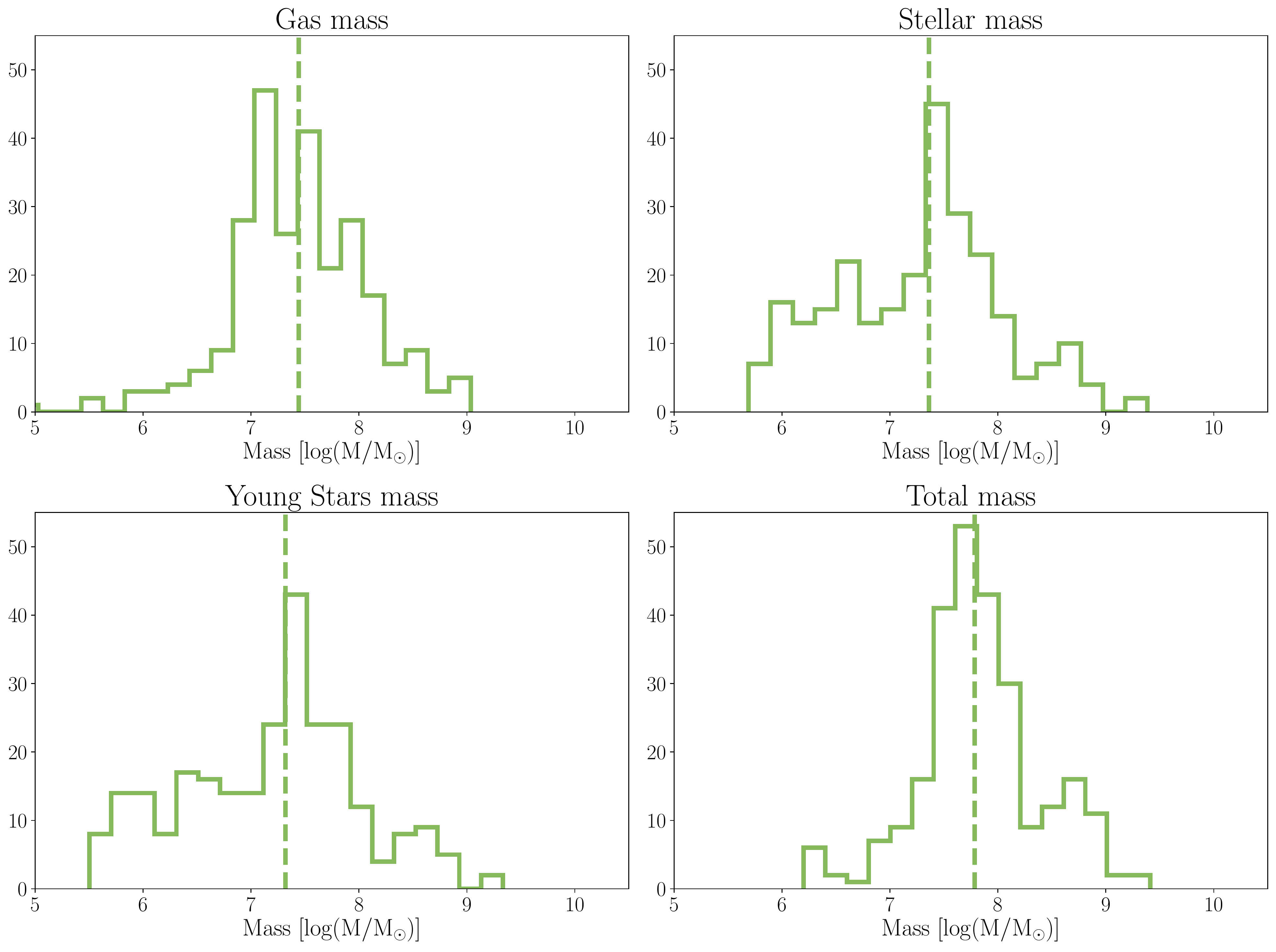}
\caption{Mass histogram of the high resolution HST-like clumps. A detail comparison simulation per simulation is shown in Figure \ref{bigplot}.}
\label{histHR}
\end{center}
\end{figure}

What we define as \textit{High resolution clumps} are the clumps found on the 12 parsec per pixel image defined in section \ref{MockObsTh}. The clumps found are shown in Figure \ref{mockHR}. Figure \ref{histHR} shows the mass histogram of all the clumps found in all four simulations. The median gas mass is $2.8\times10^7$ \msun. The median stellar masses is $3.3\times10^7$ \msun. Around 85\% of the stellar mass is made-up of stars younger than 100Myr. The median total masses is $6.1\times10^7$\msun. For more detailed mass distributions, see Appendix \ref{bigplot}.

The structures detected in those images are much less massive and more numerous than in the previous mocks thus raising the question of their relation. We answer this question in the following section.

\section{Hierarchization of the different clumps}

\subsection{Structures in the HST giant clumps}

\begin{figure}
\begin{center}
\includegraphics[width=\columnwidth]{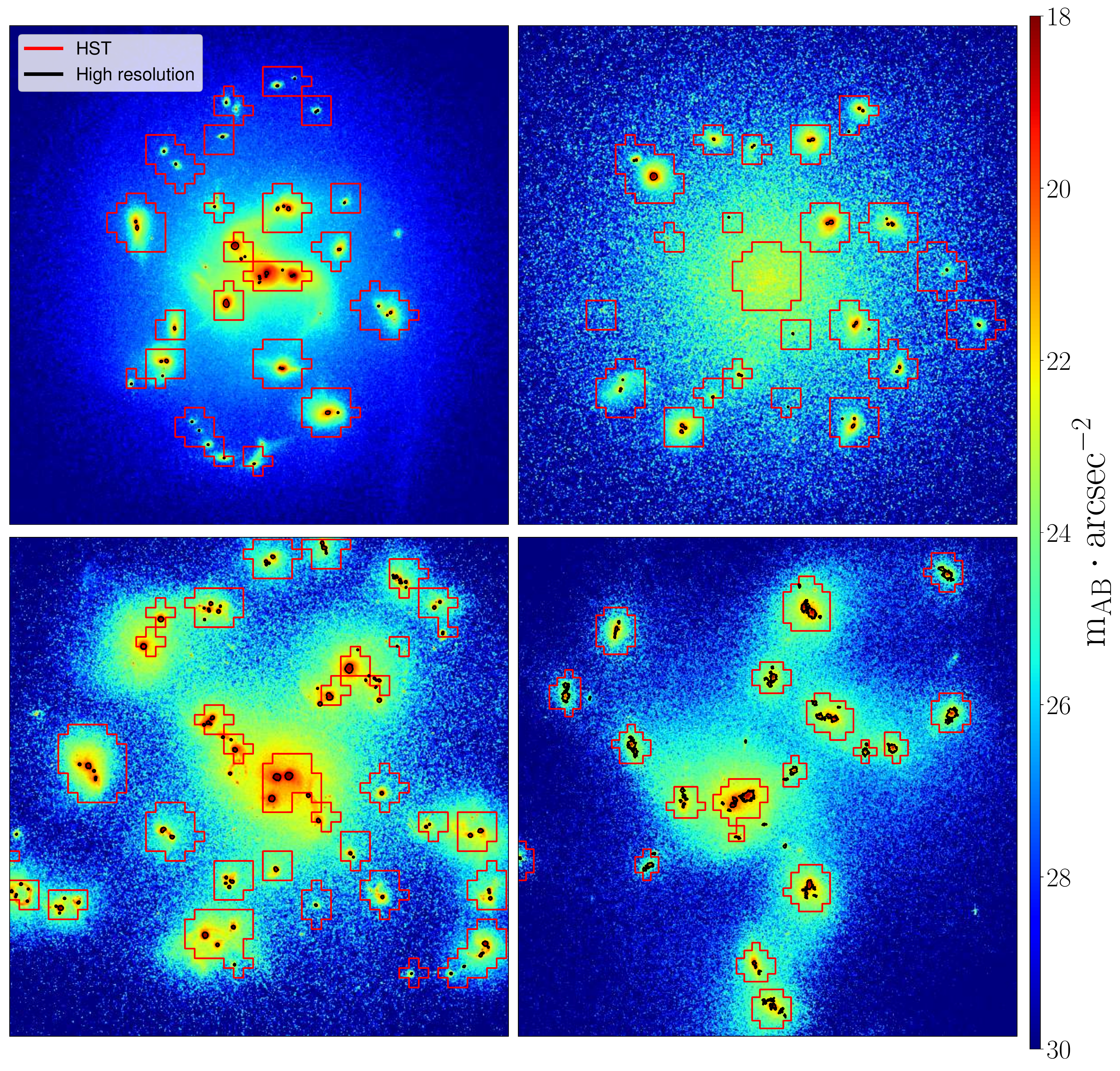}
\caption{Hierarchy of the structures in all Simulations shown over the HST like high resolution image. In red are the HST like clumps and in black the high resolution HST like giant clumps.}
\label{HierarchyAll}
\end{center}
\end{figure}

In the previous section, we showed that in our simulations clumps and structures can be detected at different scales, from the few hundred parsec scale to the ten parsec scale. This section will present how all those structures at different scales are related. Fig. \ref{HierarchyAll} presents all the different structures detected in the four non-lensed mocks superimposed over the high-resolution HST like image. One can see that all High Resolution clumps are contained inside the giant clumps. 

\begin{figure}
\begin{center}
\includegraphics[width=\columnwidth]{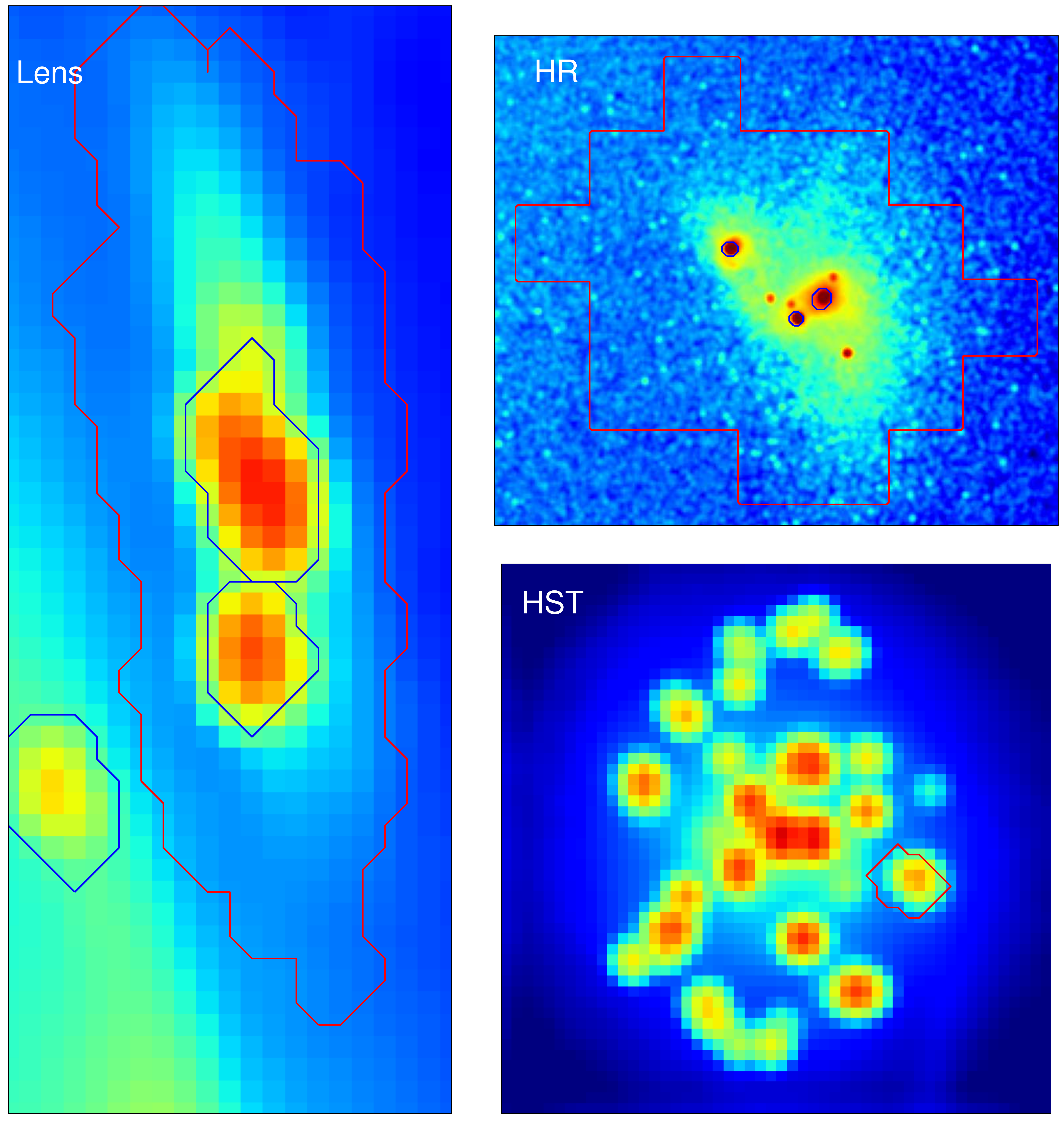}
\caption{Zoom-in of a giant UV clumps in the two 
other different mock observations. The red contour is the giant clump detected in the HST mock. The blue contours are the clumps detected in the shown mock.}
\label{ZoomClump1}
\end{center}
\end{figure}

Fig \ref{ZoomClump1} displays a giant clump of simulation 1 within the red contour in the top right panel as well as all the equivalent regions in the three mocks. One can see on the top right panel, which corresponds to the high resolution HST-like mock, bright peaks that are detected with Astrodendro. The giant clump constitutes of several high-resolution and less massive clumps. This reasoning applies to almost every giant clump as more than 60\% have two substructures or more.

The giant clumps as detected at un-lensed HST resolution contain 40-50\% of the total gas mass for roughly 25\% of the disc surface. They also contain between 100 and 90 \% of the high resolution structures and 100 and 98\% of their masses. In this sense most of the star forming small structures of the disc are located inside the giant clumps.

\subsection{Giant clumps as gravitationally bound structures}
\label{sec:virial}

In the previous subsection, we have demonstrated that giant clumps in our simulations are a superposition of several less massive substructures. This does not imply that lower-mass substructures reside in the same giant clump forever. It is known that giant clumps can lose a large fraction of their initial content, while maintaining their mass by the accretion of new material \citep[e.g.,][]{Dekel13, Bournaud14}. This can be true in particular for sub-clumps, which can leave their initial giant clump and be later-on re-accreted onto another giant clump, if they are long-lived enough \citep[as is the case, for instance, in][]{Behrendt16}. Alternatively, the sub-clumps could be short-lived and rapidly destroyed by feedback within ~10 Myr, hence being only transient sub-structures inside longer-lives giant clumps \citep[as in the case in][]{Bournaud14}.
The next question that arises is the following: are the giant clumps detected only as a random superposition of structures, like a transient chance structure or are they physically bound structures? To answer this question we compute the virial parameter of the giant clumps. This parameter is described in equation \ref{virial} and represents the competition between kinetic and gravitational energy \citep[][Equation (2.8a)]{Bertoldi92}. A value below unity means the structure is gravitationally bound.

\begin{equation}
\label{virial}
\alpha = \frac{5\sigma_v^2R_{1/2}}{GM} 
\end{equation}

\begin{figure}
\begin{center}
\includegraphics[width=\columnwidth]{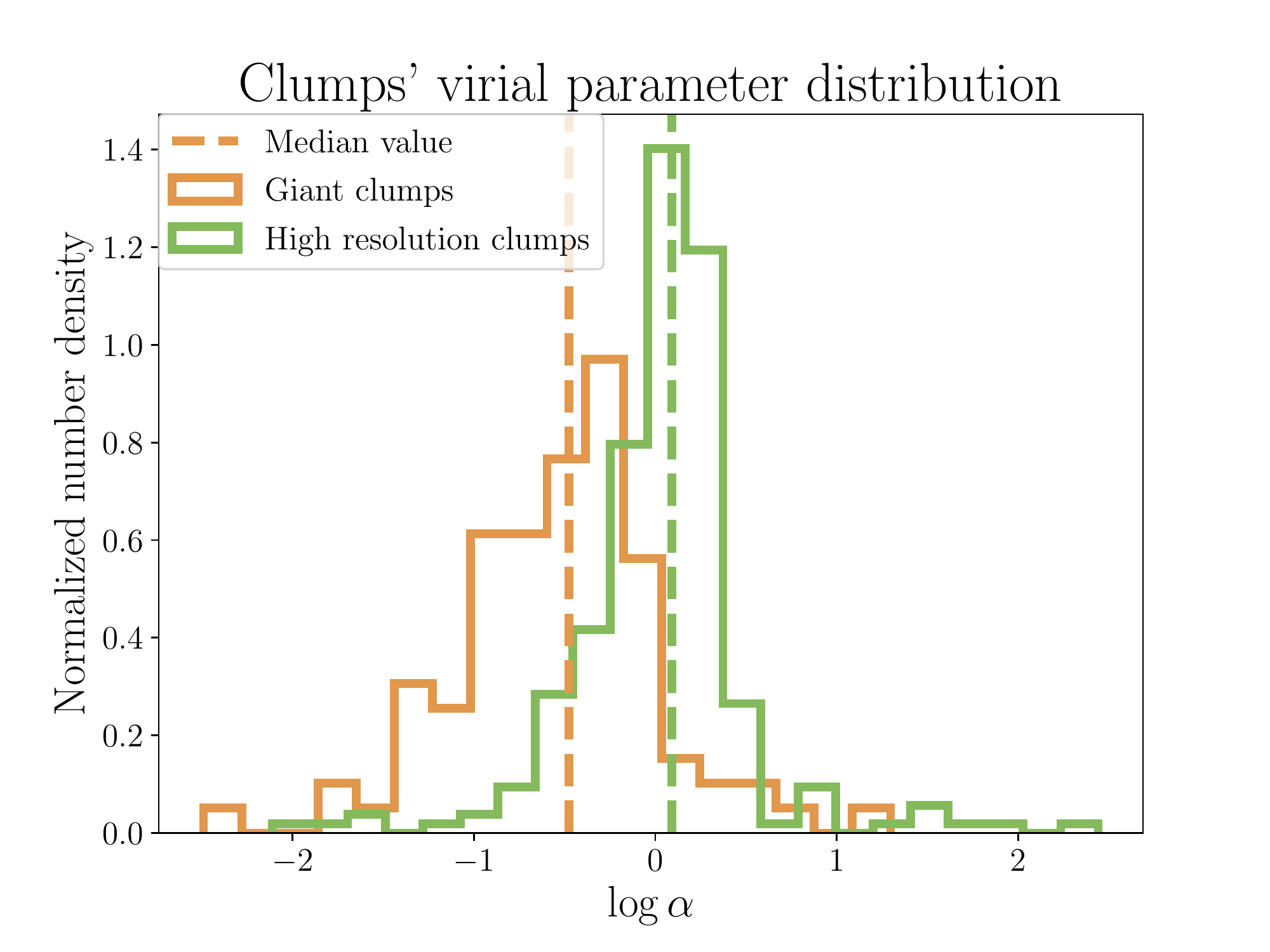}
\caption{Virial parameters of the giant clumps (orange) and the substructures (green).}
\label{alpha}
\end{center}
\end{figure}

The velocity dispersion is computed as the quadratic mean of the speed of sound and the turbulent velocity of the gas, which are computed for gas within 1kpc from the disc mid-plane in order not to be contaminated by outflows and inflows further above and below the disk plane. Rotation is not removed as attempts confirmed its negligible effect compared to turbulent motions.
Both stellar and gas mass are taken into account for the mass term in Equation \ref{virial}.
Figure \ref{alpha} shows the results for both giant clumps and high resolution structures. We can here see that most giant clumps in our simulations are gravitationally bound, with a median virial parameter of 0.33 meaning that the giant clumps are gravitationally bound structures that are unstable. Those structures are still undergoing gravitationnal instability along with their on-going collpase, and should thus be fragmenting into smaller sub-structures, which is observed to be the case. This means that as small clumps are most often gathered together into giant clumps, these giant clumps are not random chance superpositions of smaller clumps. Indeed, the giant clumps are bound structures consisting of small sub-clumps and the diffuse gas between those sub-clumps.
The stars newly formed in the gas should also be bound to the structure as they have the velocity of the gas they are formed with. \cite{Bournaud14} shows that old stars are also gravitationally bound to the clump structures.
Now that the content of the giant clumps and their physical existence has been detailed, we will discuss their formation in the next section with the understanding of our simulation.

\section{Discussion}

\subsection{Effect of feedback on structures at various scales}
\label{sec:feedbackscales}

\begin{figure}
\begin{center}
\includegraphics[width=\columnwidth]{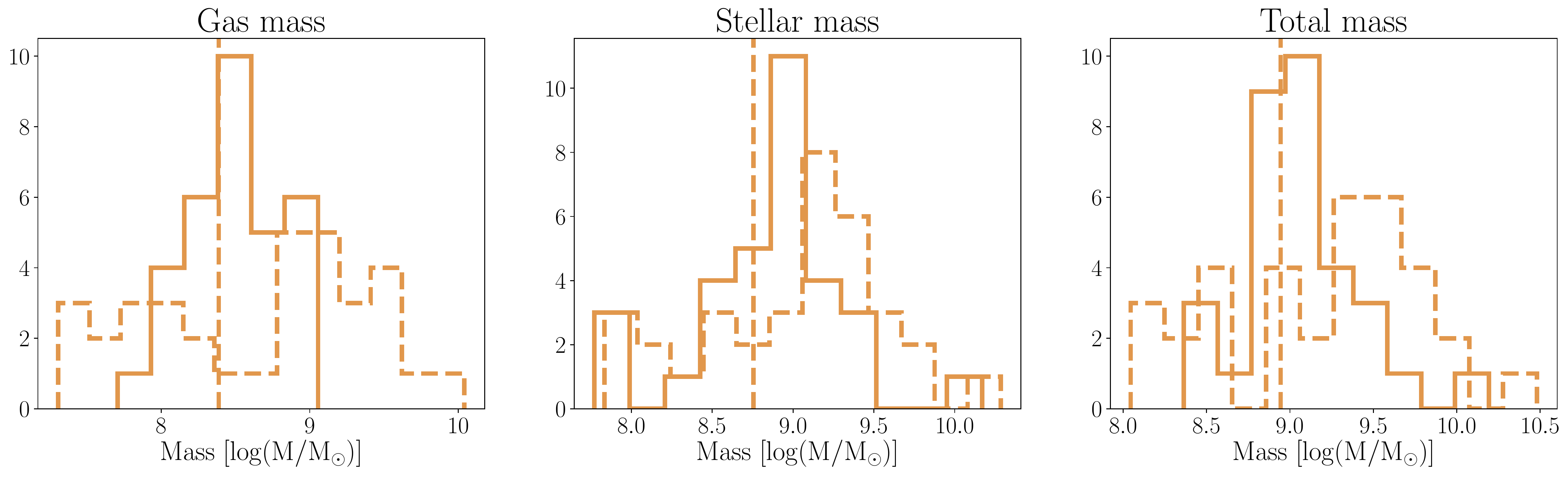}
\caption{Comparison of the HST clumps in Simulation 3 with feedback (plain lines) and without feedback (dotted lines).}
\label{lrnofb}
\end{center}
\end{figure}

\begin{figure}
\begin{center}
\includegraphics[width=\columnwidth]{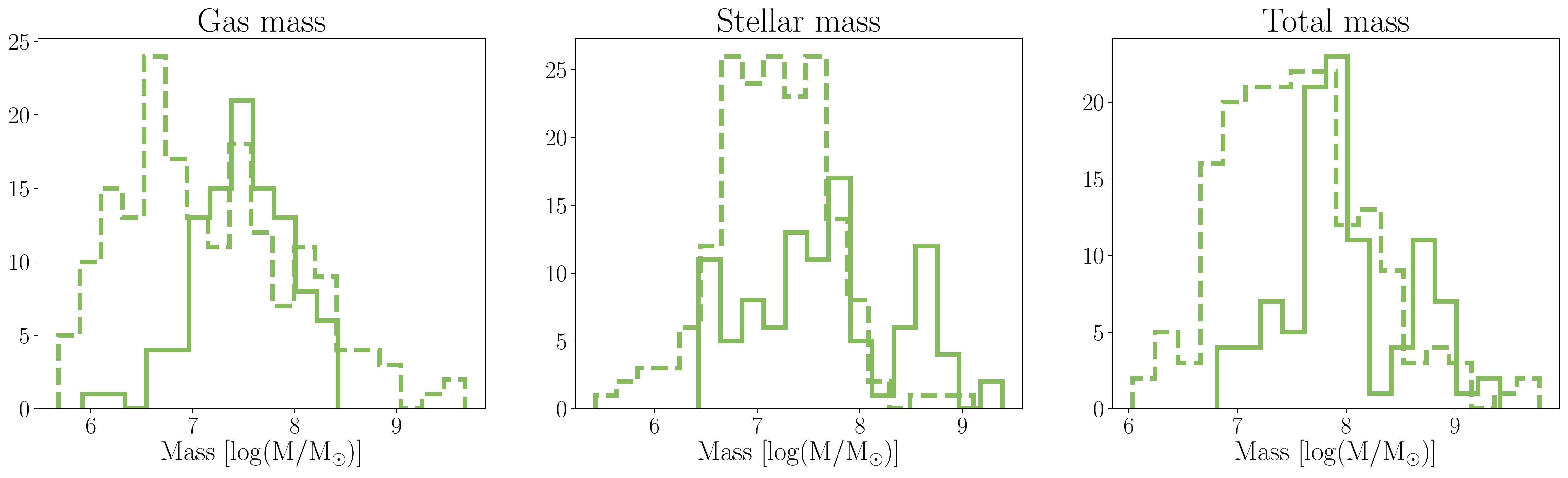}
\caption{Comparison of the high resolution HST clumps in Simulation 3 with feedback (plain lines) and without feedback (dotted lines).}
\label{hrnofb}
\end{center}
\end{figure}

Previously we have shown that most of the giant clumps are gravitationally bound in our simulations. The low mass structures have much higher virial parameter as seen on Figure \ref{alpha}. Such high virial parameter values mean that those sub-structures could have been impacted by star-formation feedback and are being destroyed : feedback expels the gas and the stars are being dispersed because of the subsequent decrease in gravitational potential. Such a scenario has been proposed by \cite{Parmentier05}, where the removal of gas in gas-rich stellar cluster could lead to the evaporation of the cluster. However, if the gas fraction of the substructures is below 50\%, the stellar sub-clump could survive and become a globular cluster, as proposed by \cite{Krumholz10} and \cite{Shapiro10}. In order to test this hypothesis we restarted simulation 3 without any feedback. After applying the exact same methods as above, Figures \ref{lrnofb} and \ref{hrnofb} show the comparison of the masses of the clumps found in the three mock observations for respectively the giant clumps and the sub-clumps. One can see the effect of the suppression of feedback: the mass range for all detected clumps broadens in particular towards the masses below $10^6$ \msun for sub clumps. This can be explained by the fact that without feedback the less massive clumps can survive for a long time while feedback quickly destroy them, which is consistent with what was discussed in Section \ref{sec:virial}. For the larger structures gas can be gradually expelled by the feedback, consistently with \cite{Bournaud14} and \cite{Dekel13}. Then, without feedback, the mass of the giant clumps can be larger than without, as seen on Figure \ref{hrnofb}. However the effect of feedback is secondary on giant clumps' properties, its impact is much more measurable on the small substructures.

\subsection{Structure formation}
\label{sec:structuresformation}

\begin{figure*}
\begin{center}
\includegraphics[width=\textwidth]{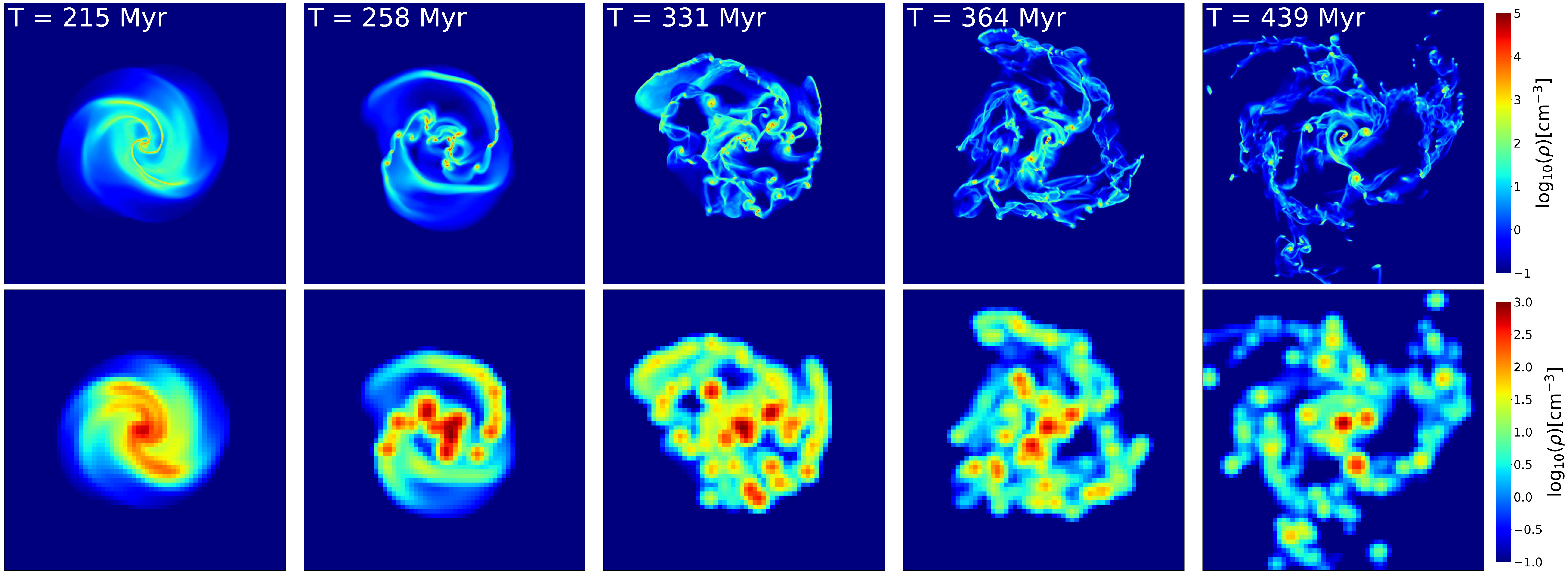}
\caption{Gas density evolution of a galaxy forming clump in the top-down scenario. The top row is the gas at a resolution equivalent to the "high resolution" case as the bottom row is at the "HST" resolution.
The first structures to collapse are the giant clumps that fragment into smaller substructures. The resolution of the images and the colorbar does not necessarily enable all sub-clumps to be visible.}
\label{topdown}
\end{center} 
\end{figure*}

\begin{figure*}
\begin{center}
\includegraphics[width=\textwidth]{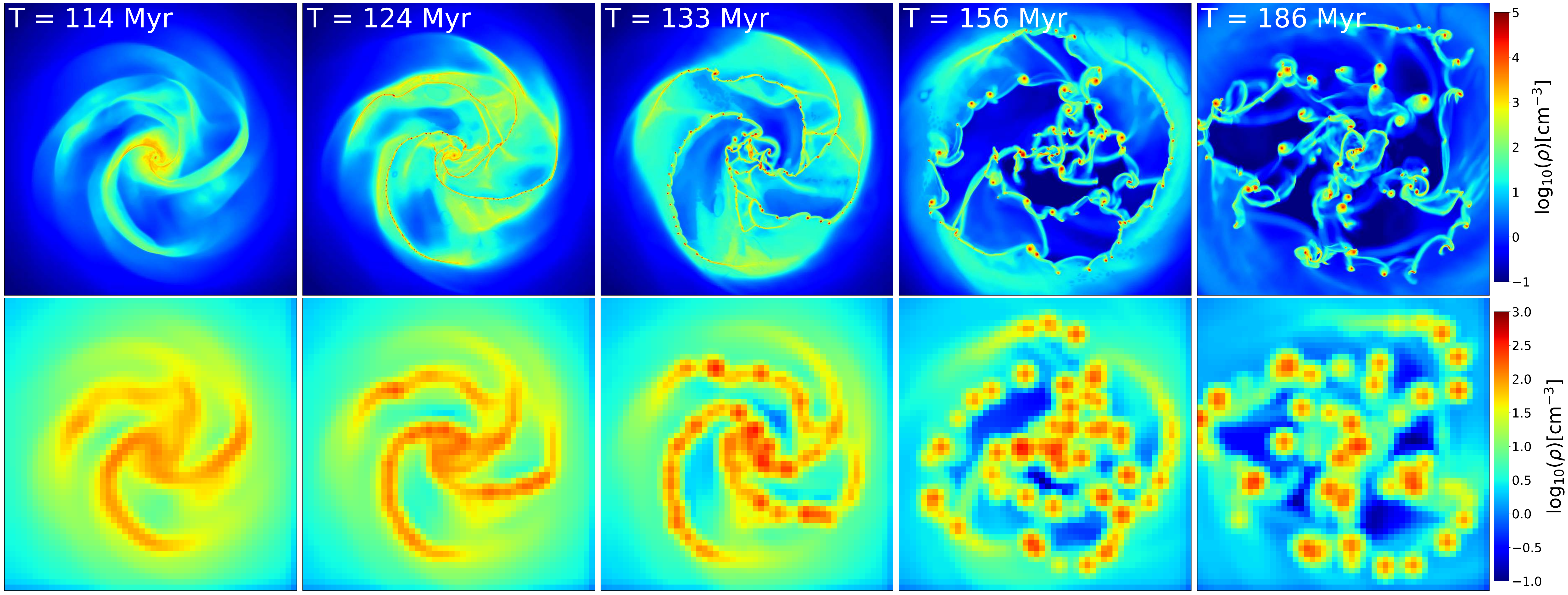}
\caption{Gas density evolution of a galaxy forming clump in the bottom-up scenario. The top row is the gas at a resolution equivalent to the "high resolution" case as the bottom row is at the "HST" resolution.
The first structures to collapse are small structures that agglomerate into larger structures: the giant clumps. The resolution of the images and the colorbar does not necessarily enable all sub-clumps to be visible.}
\label{bottomup}
\end{center} 
\end{figure*}

There is no consensus on the formation of giant clumps and sub-structures yet. Two different scenarios are currently at stake and are dependent on the initial condition of the disc, they are called \textit{top-down} and \textit{bottom-up}. In the \textit{top-down} scenario the giant clumps form first before forming sub clumps (see Fig. \ref{topdown}) as in the \textit{bottom-up} the sub clumps form first and then agglomerate to form the giant clumps \citep{Behrendt15} (see Fig. \ref{bottomup} of this paper). Simulation 1 and 2 were run to have a top-down scenario and the other two were run to have a bottom-up scenario. Those scenarios can happen only if the disc's Toomre parameter is below unity for a two-dimensional disc, or below about 0.7-0.75 for a finite thickness disc \citep[e.g.,][]{Kim02}. The Toomre parameter is defined in \cite{Toomre64} as the opposition between, on the one hand stabilisation by rotation and pression and on the other hand collapse by gravitational attraction, as in Equation \ref{toomre} where $\kappa$ is the epicyclic frequency \citep[p.165]{Binney},  $\sigma_v$ the turbulent velocity dispersion, $c_s$ the speed of sound and $\Sigma$ the surface density of the disc.

\begin{equation}
\label{toomre}
Q = \frac{\kappa\sqrt{\sigma_v^2+c_s^2}}{\pi G \Sigma}
\end{equation}

\begin{figure}
\begin{center}
\includegraphics[width=\columnwidth]{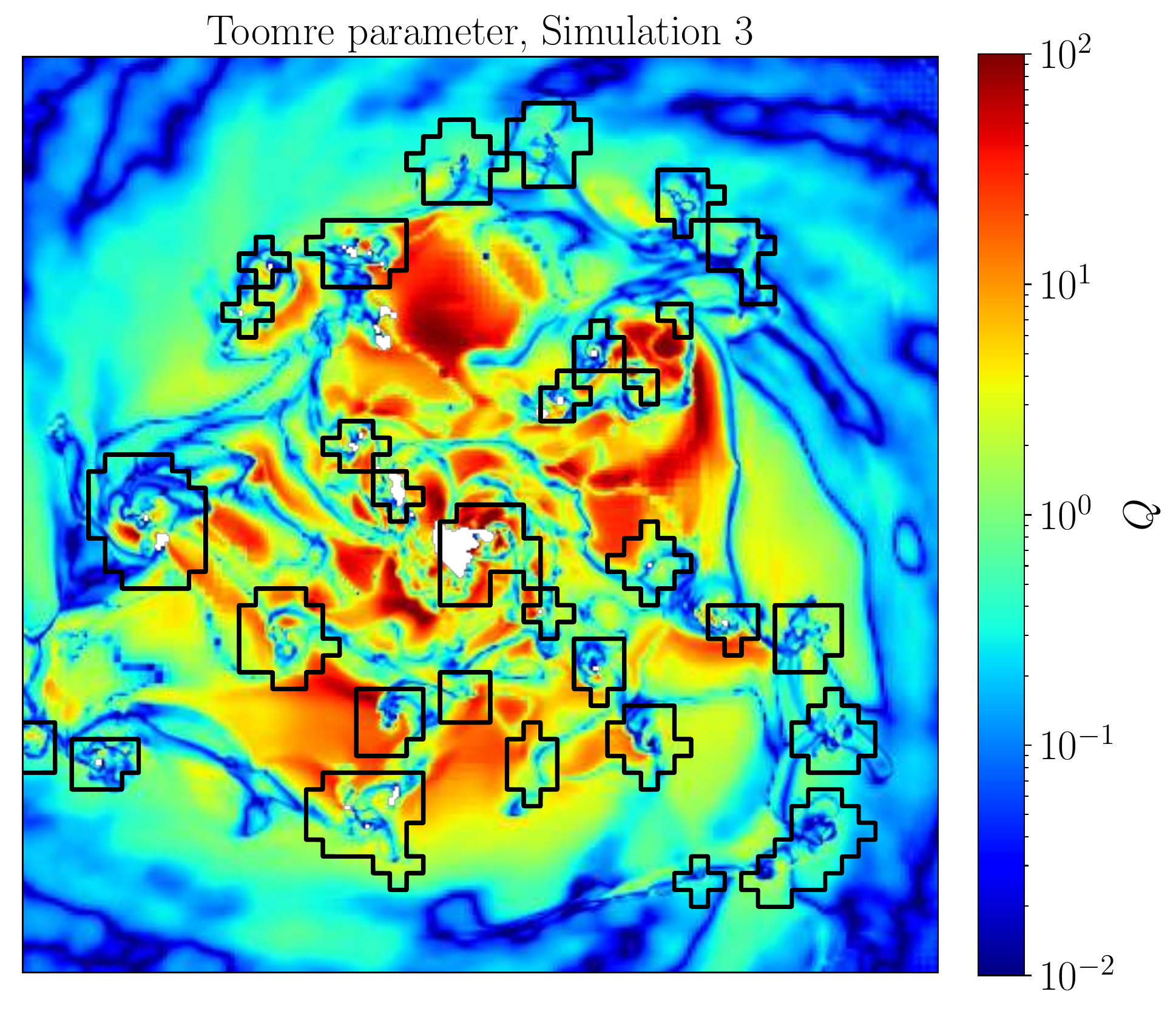}
\caption{Toomre parameter map of simulation 3. In black contour are shown the clumps detected on the HST mock observation.}
\label{toomremap}
\end{center}
\end{figure}

\begin{figure}
\begin{center}
\includegraphics[width=\columnwidth]{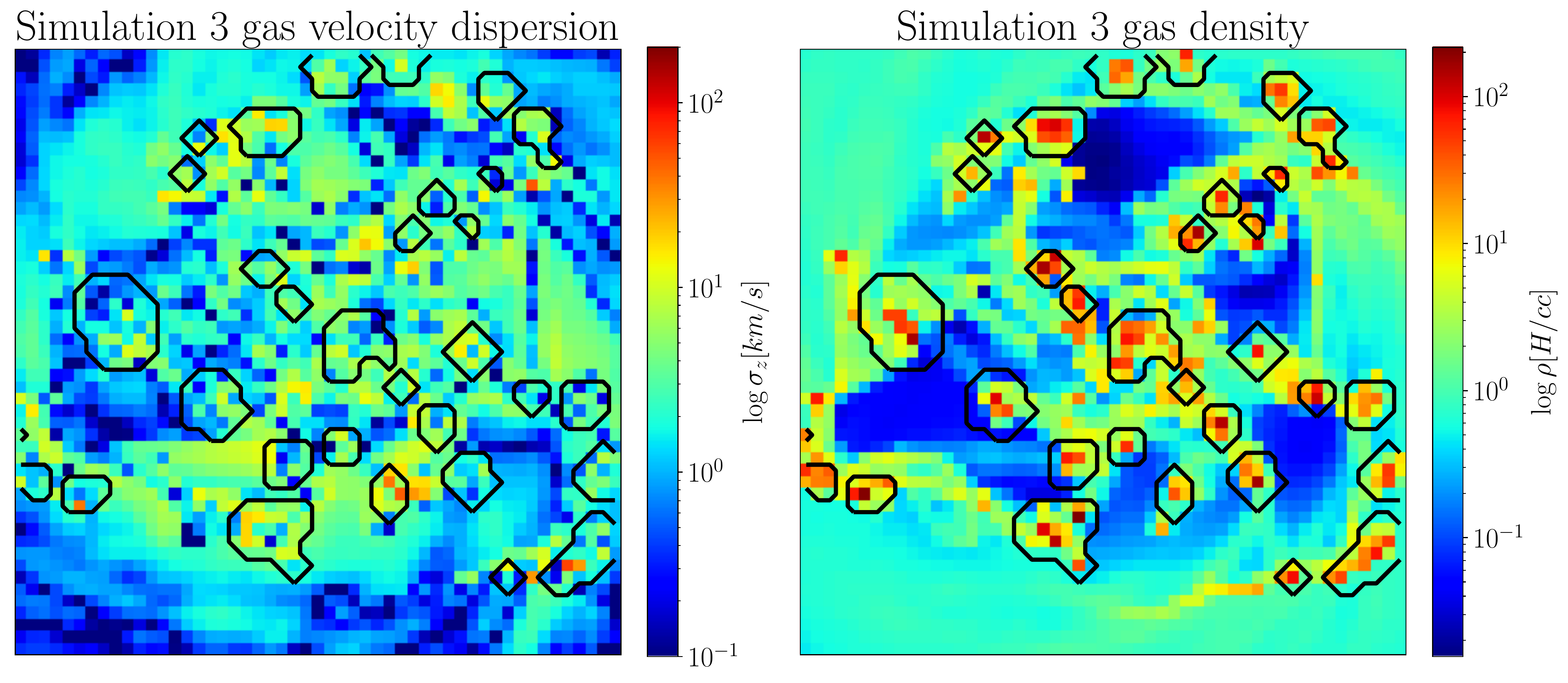}
\caption{Maps of the velocity dispersion and gas density of Simulation 3 integrated along the line of sight. The black contours correspond to the giant clumps.}
\label{sigmaz3}
\end{center}
\end{figure}

A value of $Q$ above unity for an axisymetric disc means it is stabilised by rotation and/or pressure while a value below one means it can gravitationally collapse. This is the case in our simulations where we find a value of $Q$ below unity for the giant clumps, as depicted in Figure \ref{toomremap}: they are regions that could have collapsed gravitationally when the disc was axisymetric. One can try to estimate the Toomre parameter of a proto-clump region, i.e. before it collapses into a giant clump, based on the fact that the only physical parameter in Equation \ref{toomre} that largely varies during the collapse is the surface density that is multiplied by a factor about 10. The velocity dispersion is not found to increase in our simulations, especially inside giant clumps. This can be seen in Figure \ref{sigmaz3} where we can visualise that the turbulent velocity dispersion in each clump does not differ from the one outside and is not dependent of the gas density. Furthermore, the speed of sound does not significantly impact the quadratic mean in Equation \ref{toomre} even if the clump is inside an HII region. Indeed with a temperature of an HII region of $2.5 \times 10
^4 \mathrm{K}$, the speed of sound will be close to 10km/s, which is largely dominated by the turbulent speed that is roughly around 30-50 km/s.
The epicyclic frequency does not substantially vary after the collapse of the clumps as it is only radius dependent. This means that between a proto-clump and a clump the Toomre parameter can decrease by at most a factor about 10. As in our simulations the Toomre parameter inside collapsed regions is below 0.1, the value before the collapse could not have exceeded unity meaning collapse of giant clumps can be due to gravitational instability.

We just showed that from the analysis of the disc, our simulations are compatible with formation of structures through violent disc instabilities \citep{DSC09}. This is true for all our simulations, regardless of the scenario of clump formation we imposed. In addition the clumps formed have similar properties in previous studies \citep{Ceverino15, Behrendt16} where the simulated galaxies also have similar properties. Clumps are also found in disks with similar dynamics in \cite{Leung20}, where the clumps are not high peaks of velocity dispersion. However, they are smaller in size and mass compared to our giant clumps and do not seem to be gravitationally bound. Those clumps are therefore closer to the substructures we find in our giant clumps.

The gravitational collapse of the disc then occurs at a mass above the Jeans' mass \citep{Jeans1902}, which is defined as the equilibrium between the pressure and the gravitational force. It is defined in Equation \ref{jeansmass} where $\sigma_v$ is the velocity dispersion and $c_s$ the speed of sound.

\begin{equation}
\label{jeansmass}
\mathrm{M}_\mathrm{J} = \frac{\pi^{\frac{5}{2}}}{6} \frac{\left(\sigma_v^2+c_s^2\right)^\frac{3}{2}}{G^\frac{3}{2}\rho^\frac{1}{2}}
\end{equation}

As the Jeans' mass depends on the speed of sound, it also depends on the gas temperature. On one hand, the cooler the gas, the lower the Jeans' mass thus formation of smaller structures. On the other hand if the temperature is higher, the Jeans' mass will get higher.
The simulations from this paper were run with an initial high temperature (Simulation 1 and 2) and with an initial cold disc (Simulation 3). The one from Behrendt et al. (in prep) starts also with cold initial disc.

In the case of an initial hot disc its Jeans' mass is high and leads to the formation of large structures that cool down, lowering their Jeans' mass and allowing them to fragment into smaller substructures: the \textit{top-down} scenario. This scenario is observed in Simulation 1 and 2.

For the simulation 3, the way cooling is implemented forces the disc to be initially cold ($10^4$ K) which leads to a Jeans' mass lower than in the Simulation 1 and 2. As the disc's Jeans' mass is lower the first structures to form are less massive than the giant clumps. Those structures stir the surrounding gas, increasing the turbulence and thus the velocity dispersion, leading to an increase of the Jeans' mass of the disc. A second collapse is then happening at a higher mass: the giant clumps are formed and capture the first structures that become the sub-clumps observed at high resolution. This is the \textit{bottom-up} scenario.

Remarkably we do not find any evidence of those different scenarios if we compare the physical properties of the giant clumps and their sub-clumps through the different simulations. One could say that in the \textit{top-down} scenario, as the sub-clumps are formed by the fragmentation of the giant clumps, their could be a link between the Jeans' mass of the giant clumps and the mass of it sub-clumps. Conversely, in the \textit{bottom-up} scenario there would be no link between those two values. 
One could also argue that the different formation history could lead to different behavior of the inter-sub-clumps medium inside giant clumps that could be probed by the velocity dispersion of those regions.
The values we compared between the different scenarios are the following:
\begin{itemize}
\item Masses of gas and stars in giant clumps,
\item Masses of gas and stars in sub-clumps,
\item Jeans' masses at the scale of the giant clumps,
\item Jeans' masses at the scale of the sub-clumps,
\item 1D-velocity dispersion at every scale,
\item 1D-velocity dispersion ratio between giant clumps and the whole galaxy,
\item 1D-velocity dispersion ratio between giant clumps and their sub-clumps,
\item 1D-velocity dispersion ratio between sub-clumps and their surrounding gas.
\end{itemize}
The comparison of all these physical parameters did not show any statistically significant difference between the clumps formed \textit{bottom-up} and those \textit{top-down}. All of those arguments make us think that the initial formation scenario of the giant clumps is not relevant to understand their evolution and that they tend to become very similar structures however they are formed.

\subsection{Detectability with ALMA}

\label{sec:ALMAthreshold}

In this paper we focused mainly on resolution effect on the clumps detection. Nevertheless in order to link observations and simulations one needs to understand sensitivity effect as clumps are not detected in recent observations with ALMA \citep{Cibinel17,Rujopakarn19,Ivison20}. We created mock observations with ALMA OST as described in Section \ref{sec:ALMAOST}.

\begin{figure*}
\begin{center}
\includegraphics[width=\textwidth]{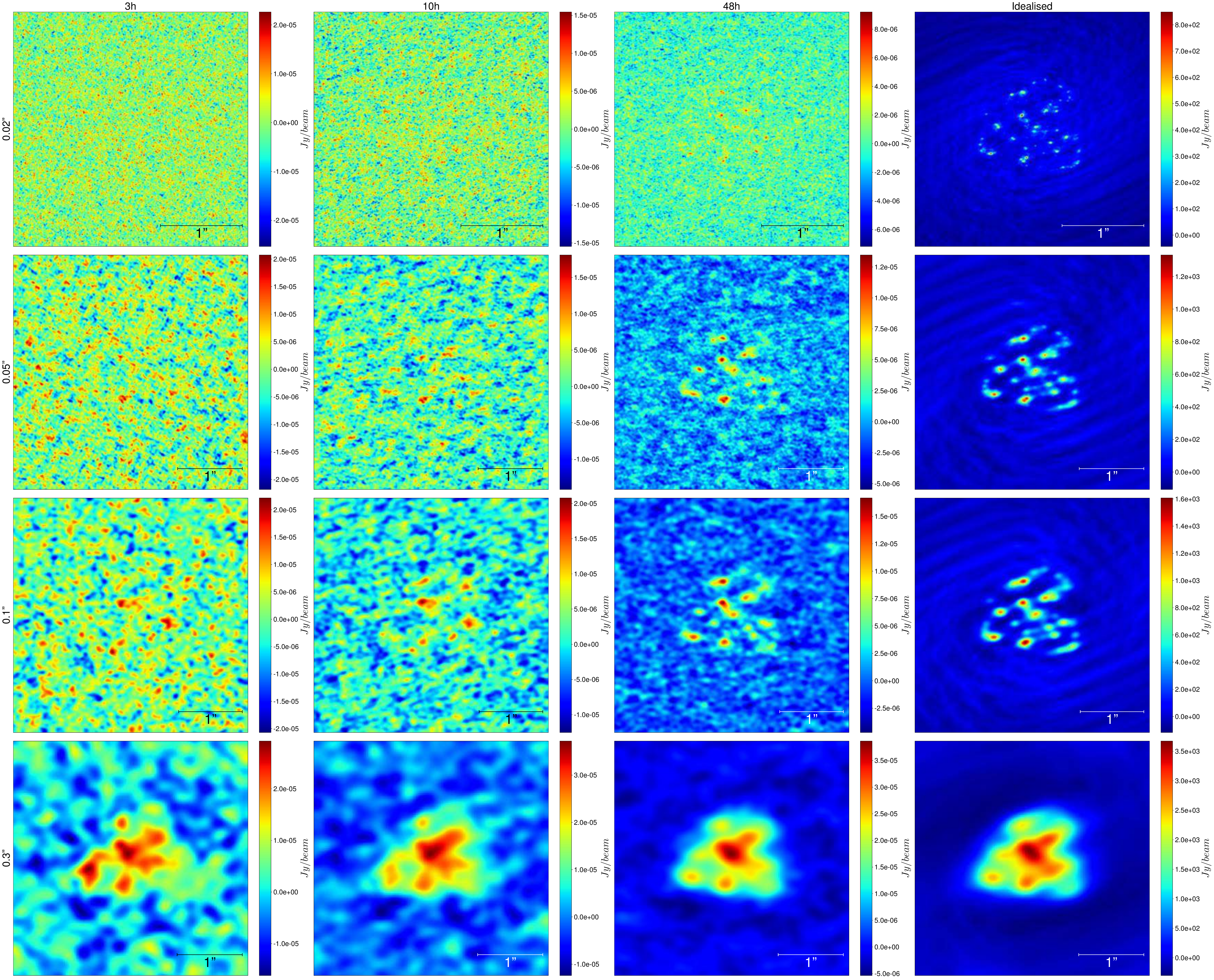}
\caption{Mock observation created with the ALMA Observation Support Tool. From the top to the bottom the beam size is increasing, and the observation time is increasing from the left to the right. The last column shows the idealised observation.}
\label{ostplot}
\end{center}
\end{figure*}

The mocks for simulation 3 are represented in Figure \ref{ostplot}. A long observation time of more than 48h is needed to detect structures with a beam size of 0.02 arcsecond, corresponding to a physical size of 170 parsec at z = 2. In the case of an observation time of 10h with the 0.02 arcsecond beam, some subclumps can be detected but most of them are dominated by noise. With a larger beam the clumps can be detected in around ten hours and some substructures are at the verge of the detection. This first statement is qualitative and a more quantitative case follows later in the section.

\begin{figure}
\begin{center}
\includegraphics[width=\columnwidth]{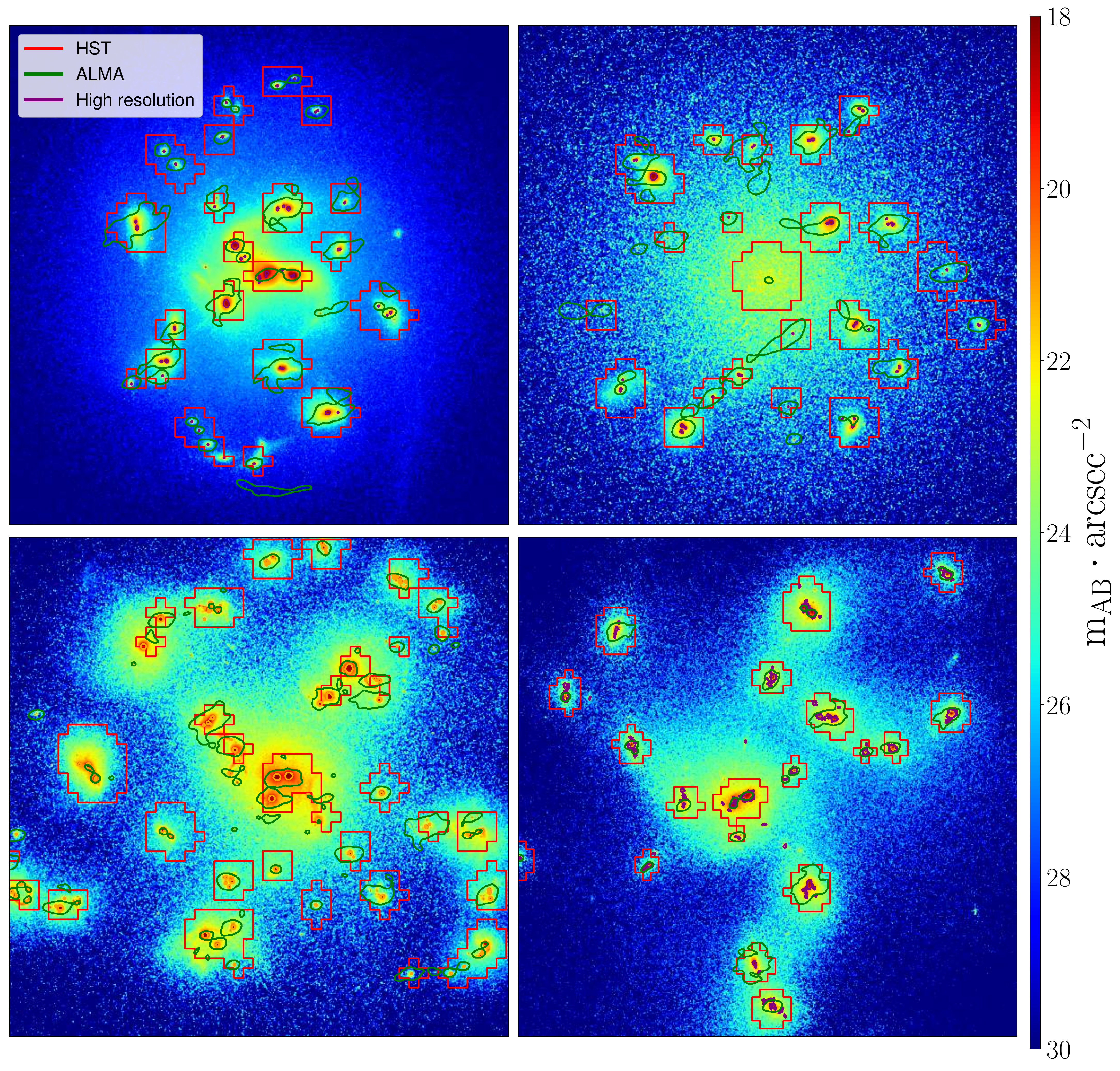}
\caption{Hierarchy of the structures in all Simulations shown over the HST like high resolution image. In red are the HST like clumps, in purple the high resolution HST like giant clumps and in green the ALMA clumps at a 0.02 arcseconds resolution.}
\label{HierarchyAlma}
\end{center}
\end{figure}

\begin{figure}
\begin{center}
\includegraphics[width=\columnwidth]{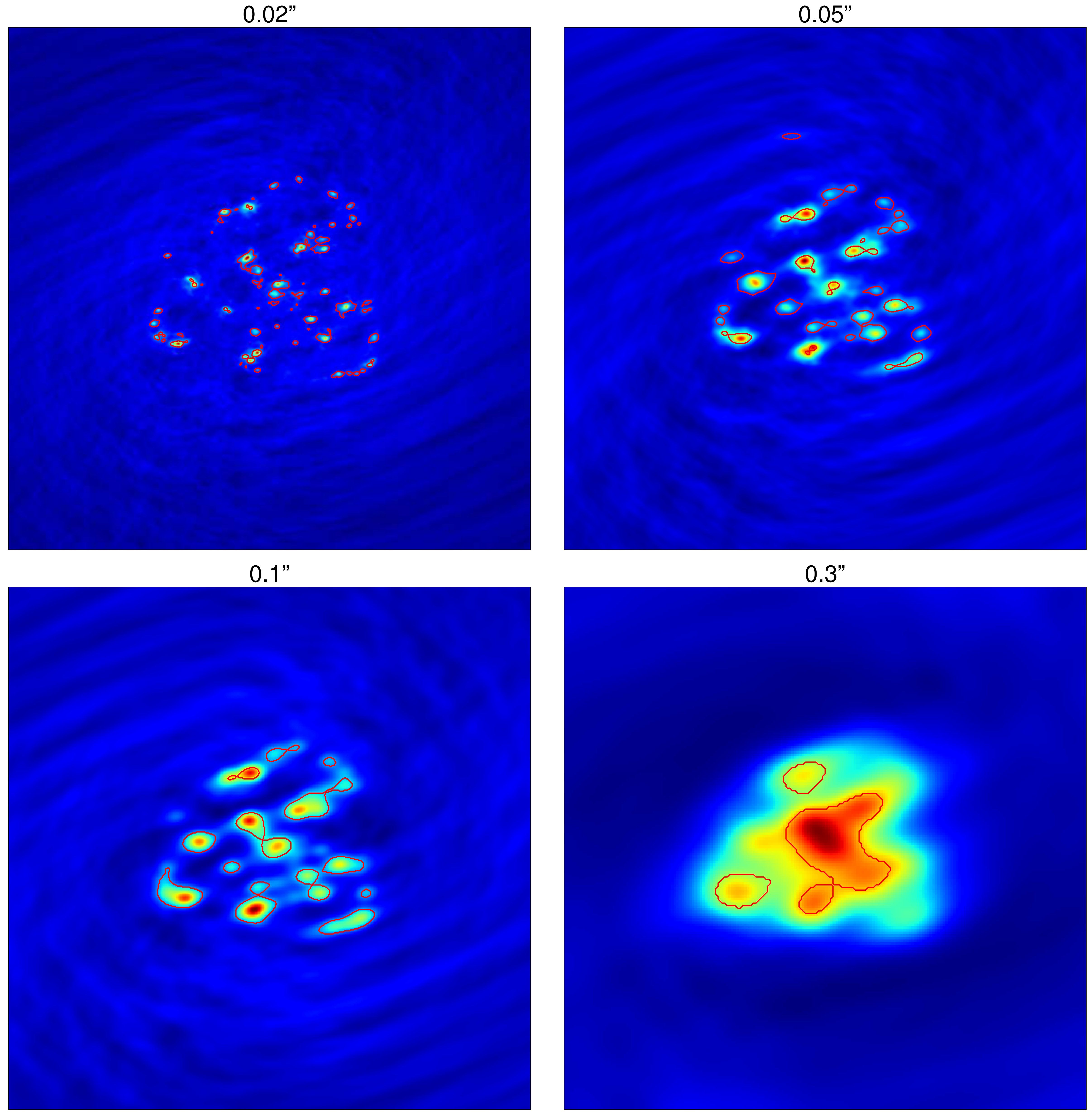}
\caption{Clumps detected in the idealised mock of Simulation 3, with negligible noise and extremely high sensitivity, ALMA simulations made with OST with different beam sizes. From left to right and top to bottom: 0.02", 0.05", 0.1", 0.3".}
\label{ostclumps}
\end{center}
\end{figure}

By running Astrodendro on the idealized mocks only we detect structures that are represented in Figure \ref{ostclumps}. A comparison of the clumps found on the most resolved ALMA mocks with the one found on the mocks presented earlier is shown in Figure \ref{HierarchyAlma}. One can see that at this resolution ALMA does not resolve the giant clumps (in red on Figure \ref{HierarchyAlma}). For a fraction of those only the center of the clump is resolved even if the outlying region is gravitationally bound and is part if the giant clump, leading to a possible underestimation of the mass. For the remaining fraction the ALMA clumps correspond to the sub-structures (in purple on Figure \ref{HierarchyAlma}) therefore could probe the inner structure of the giant clumps.

\begin{figure*}
\begin{center}
\includegraphics[width=\textwidth]{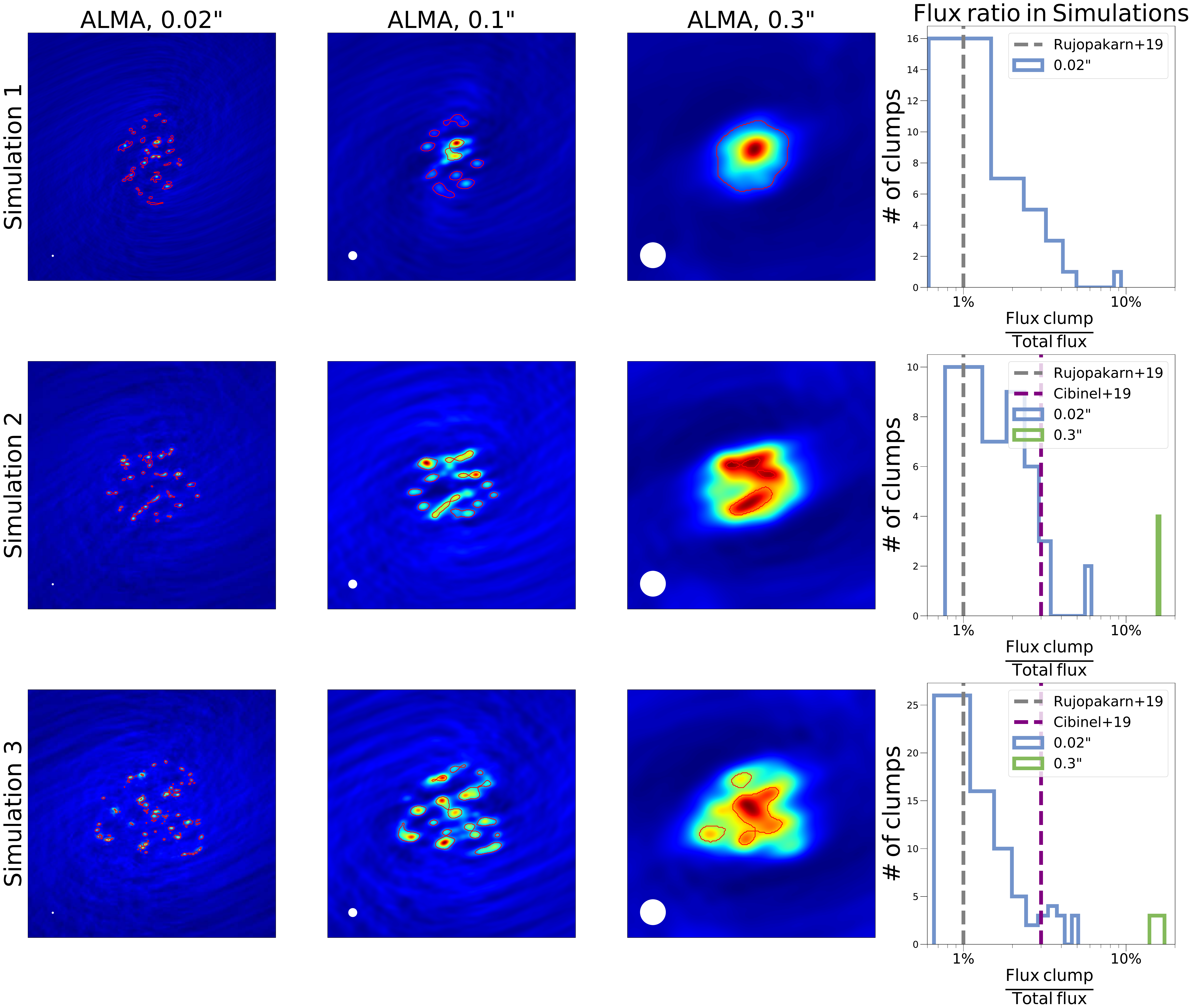}
\caption{Left column is the idealized, with negligible noise and extremely high sensitivity, ALMA mock made with OST with a 0.02" beam. The second column is the one with a 0.1" beam and the third column is with a 0.3" beam. Ratios between flux of the clumps detected on the previous columns and the total flux of the galaxy are on the right column. The red contours are the clumps detected with Astrodendro. White circle are the beams sizes.}
\label{ostflux}
\end{center}
\end{figure*}

We then compute the flux ratio relative to the flux of the whole galaxy of each of the detected structure on the ALMA idealised mocks excluding detection closer than 2 kpc from the center of the galaxy. The result is shown in Figure \ref{ostflux} for the three simulations. 

The comparison with previous studies, like \citet{Rujopakarn19}, whose galaxies have similar position with respect to the main sequence and where the upper limit is of 1\% with a 200-pc beam, suggests that either (1) the galaxies they observed are clumpy with lower mass clumps than in our models that are tuned to correspond to very clumpy galaxies, or (2) the sub clumps in these galaxies are gas-poor, potentially under the effect of strong stellar feedback. This suggests that our model of feedback is not strong enough to efficiently deprive sub-clumps from their gas. A stronger feedback implementation could be enough to deprive them from gas or destroy them without destroying giant clumps which is a scenario proposed in \cite{Bournaud14}. One might need to be careful as if the feedback is too strong the substructures but also the giant clumps could be destroyed as seen in \citet{Hopkins12} and \cite{Tamburello15}. Larger observational datasets of this type including very clumpy galaxies could disentangle these interpretations and potentially probe feedback effects.

For a comparison with \cite{Cibinel17} the number of detected structures is very dependent on the simulations as seen on Fig. \ref{ostflux}. The beam size of 0.3" being large, the giant clumps are blended-in together into even larger and brighter clumps as it is the case for Simulation 2 and 3. The opposite is also seen in Simulation 1 where the clumps are blended-in together without being local maximum in luminosity. The galaxy presents asymmetry but nor the clump finder nor the eyes can detect any giant clumps.
Therefore, depending on the between-clumps distance the giant clumps can be detected or not with ALMA at resolution matching \cite{Cibinel17}. The detected clumps are very luminous, with a flux ratio to total flux around 15\% as they are the result of the smearing of several giant clumps.

\subsection{Effect of dust extinction}

\begin{figure}
\begin{center}
\includegraphics[width=\columnwidth]{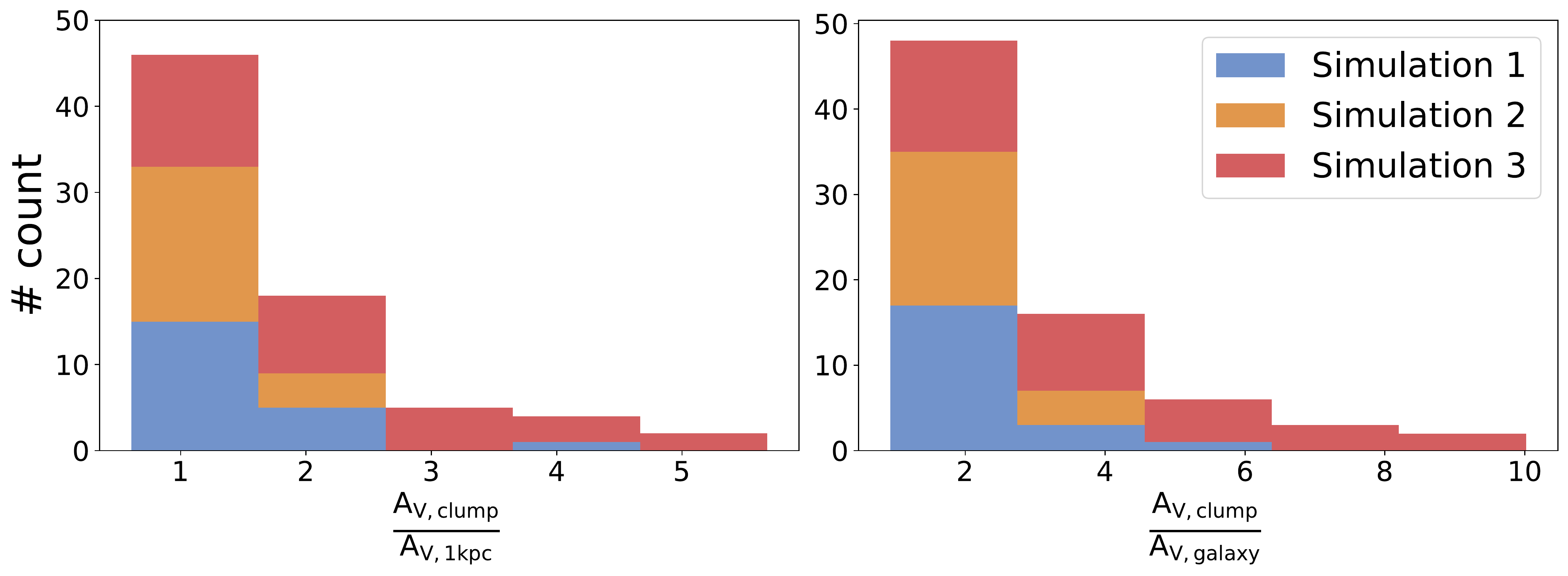}
\caption{Quantification of the dust extinction in the giant HST clumps. Left panel is the ratio of the clump dust extinction to the one of the central kiloparsec. Right panel is the ratio to the galaxy one.}
\label{Avclumps}
\end{center}
\end{figure}

In this short section we want to understand how the dust extinction will impact the detection of the giant clumps, as it was raised by \cite{Zanella21}. Indeed, they showed that for higher redshift ($z \sim 6$), dust could lead to a non detection of the internal structures of the galaxy in UV HST images.
By using the model from \cite{Guver09}, that links gas column density to $\mathrm{A_V}$, and assuming solar metallicity everywhere in the galaxy, one can estimate the dust extinction of the clumps. The comparison to the central kiloparsec and the whole galaxy is in Figure \ref{Avclumps}. One can see that clumps are not particularly dustier than the galaxy or its center and are in agreement with \cite{Elmegreen05} and \cite{Elmegreen07} which mean that real observations of such galaxies would lead to the detection of almost all the giant clumps. However, those results only show a tendency. Indeed, a much proper and deeper analysis including all physical processes that affect dust, such as feedback and radiation, would be necessary to draw a proper conclusion on the dust extinction.

\section{Conclusion}

By running four different simulations of idealized and isolated galaxies and by creating mock observations out of those in order to inspect the detection of giant clumps on resolution effects, in this paper we have shown that:

\begin{itemize}
\item The typical mass of the clumps depends strongly on the spatial resolution. They are observed starting from $10^8\//10^9\mmsun$ with HST, the so called giant clumps. The giant clumps are then not detected with higher resolution where structures have masses around $10^6\//10^7\mmsun$. Similarly, when high effective resolution is reached through strong gravitational lensing, giant clumps are not detected anymore but sub-clumps are. Those results are in agreements with \cite{Behrendt19}.
\item The smaller clumps where most of the star formation occurs at high resolution are all located inside giant clumps and not elsewhere: the giant clumps are clusters of stellar clusters.
\item Giant clumps are gravitationally bound structures meaning they have a physical existence and are not chain like structures detected due to the lack of resolution.
\item Most of the physical properties of giant clumps do not depend on their formation scenario.
\item If high resolution is used with ALMA, there will be no detection of giant clumps but only of the substructures. A much higher sensitivity will be needed to detect to structures.
\end{itemize}

Those results are obtained with four different simulations that all have different initial properties, feedback parameters, cooling modes and clump formation histories (\textit{top-down} or \textit{bottom-up}). By design our modelled galaxies correspond to the most gas-rich and clumpy ones at $\mathrm{z}\simeq2$. The same recipe used here are able to produce less clumpy galaxies if the gas fraction is lowered, with only low-mass and short-lived clouds (see for example \citealt{Renaud15}). In addition, a change in the gas fraction could change the properties of the giant clumps such as their lifetime and boundedness \citep{Oklopcic17, Fensch20}.

Our simulations are not incompatible with studies where only clumps smaller in size and mass are detected, such as \cite{Leung20, Zanella21}, whose clumps are closer to the sub-structures of the giant clumps.

However, as our giant clumps are extended and made of substructures they are widely different from the very massive ($> 10^8 \mmsun$) clumps found in \cite{Tamburello15} simulations. Indeed, their clumps are very dense and do not appear to have any structure, making them incompatible with observations of lensed galaxies.

Observations with more resolution than the typical HST one fail to identify giant stellar and gaseous clumps. Our work shows that such observations should tend to resolve giant complexes into smaller sub-clumps. The masses and sizes of sub-clumps in our models are consistent with the one detected by \cite{Dessauges17} for the stellar component and by \cite{Dessauges19} for the gaseous component.
Recent ALMA observations \citep{Rujopakarn19, Ivison20} typically employed resolution that resolve giant clumps into sub-clumps but with a sensitivity too low to detect those putative sub-clumps, which would be required to prove the gathering of sub-clumps into giant clumps.
Lower resolution ALMA observation \citep{Cibinel17} merely reach the resolution required to directly detect giant clumps according to our model.
All these observations thus remain consistent with the presence of giant clumps of stars and gas that potentially survives feedback in high-redshift galaxies. Higher sensitivity ALMA observations would be necessary to detect giant clumps along with their sub-clumps.

Finally, the hierarchy of the giant clumps revealed in this work seems to be in accordance with \cite{Fisher17}, which present local analogues of turbulent, clumpy disk galaxies. Indeed, they observe clumps which are not massive and large enough to be qualified as giant but are similar to our substructures. However, they show that by degrading the resolution, as if the galaxies are observed at a higher redshift, those structures merge into larger one that are similar to our giant clumps and thus might be gravitationally bound.

\section*{Acknowledgement}
We thank the anonymous referee for the useful comments and discussion, which greatly improved the paper. This work has been carried out thanks to the support of the ANR 3DGasFlows (ANR-17-CE31-0017). This work was granted access to the HPC resources of CINES and TGCC under the allocations 2018-A0050402192, 2019-A0070402192, and 2020-A0090402192 made by GENCI. The research of Andreas Burkert and Manuel Behrendt was supported by the Excellence Cluster ORIGINS which is funded by the Deutsche Forschungsgemeinschaft 
(DFG, German Research Foundation) under Germany's Excellence Strategy – EXC-2094 – 390783311. The simulations by Manuel Behrendt were performed on the Hydra supercomputer at the Max Planck Computing and Data Facility (MPCDF).
This research made use of Astropy,\footnote{http://www.astropy.org} a community-developed core Python package for Astronomy \citep{Astropy13, Astropy18} and astrodendro\footnote{http://www.dendrograms.org/}, a Python package to compute dendrograms of Astronomical data. Furthermore, Manuel Behrendt made use of MERA \citep{MERA}, a Julia package to efficiently load and analyze  RAMSES simulation data.

\section*{Data Availability}

The data underlying this article will be shared on reasonable request to the corresponding author.

\bibliographystyle{mnras}
\bibliography{lib}

\begin{thebibliography}{}
\makeatletter
\relax
\def\mn@urlcharsother{\let\do\@makeother \do\$\do\&\do\#\do\^\do\_\do\%\do\~}
\def\mn@doi{\begingroup\mn@urlcharsother \@ifnextchar [ {\mn@doi@}
  {\mn@doi@[]}}
\def\mn@doi@[#1]#2{\def\@tempa{#1}\ifx\@tempa\@empty \href
  {http://dx.doi.org/#2} {doi:#2}\else \href {http://dx.doi.org/#2} {#1}\fi
  \endgroup}
\def\mn@eprint#1#2{\mn@eprint@#1:#2::\@nil}
\def\mn@eprint@arXiv#1{\href {http://arxiv.org/abs/#1} {{\tt arXiv:#1}}}
\def\mn@eprint@dblp#1{\href {http://dblp.uni-trier.de/rec/bibtex/#1.xml}
  {dblp:#1}}
\def\mn@eprint@#1:#2:#3:#4\@nil{\def\@tempa {#1}\def\@tempb {#2}\def\@tempc
  {#3}\ifx \@tempc \@empty \let \@tempc \@tempb \let \@tempb \@tempa \fi \ifx
  \@tempb \@empty \def\@tempb {arXiv}\fi \@ifundefined
  {mn@eprint@\@tempb}{\@tempb:\@tempc}{\expandafter \expandafter \csname
  mn@eprint@\@tempb\endcsname \expandafter{\@tempc}}}

\bibitem[\protect\citeauthoryear{{Agertz}, {Teyssier}  \& {Moore}}{{Agertz}
  et~al.}{2009}]{Agertz09}
{Agertz} O.,  {Teyssier} R.,   {Moore} B.,  2009, \mn@doi [\mnras]
  {10.1111/j.1745-3933.2009.00685.x}, \href
  {https://ui.adsabs.harvard.edu/abs/2009MNRAS.397L..64A} {397, L64}

\bibitem[\protect\citeauthoryear{{Astropy Collaboration} et~al.,}{{Astropy
  Collaboration} et~al.}{2013}]{Astropy13}
{Astropy Collaboration} et~al., 2013, \mn@doi [\aap]
  {10.1051/0004-6361/201322068}, \href
  {https://ui.adsabs.harvard.edu/abs/2013A&A...558A..33A} {558, A33}

\bibitem[\protect\citeauthoryear{{Astropy Collaboration} et~al.,}{{Astropy
  Collaboration} et~al.}{2018}]{Astropy18}
{Astropy Collaboration} et~al., 2018, \mn@doi [\aj] {10.3847/1538-3881/aabc4f},
  \href {https://ui.adsabs.harvard.edu/abs/2018AJ....156..123A} {156, 123}

\bibitem[\protect\citeauthoryear{{Behrendt}}{{Behrendt}}{2020}]{MERA}
{Behrendt} M.,  2020, ManuelBehrendt/Mera.jl, \mn@doi{10.5281/zenodo.3675255},
  \url {https://doi.org/10.5281/zenodo.3675255}

\bibitem[\protect\citeauthoryear{{Behrendt}, {Burkert}  \&
  {Schartmann}}{{Behrendt} et~al.}{2015}]{Behrendt15}
{Behrendt} M.,  {Burkert} A.,   {Schartmann} M.,  2015, \mn@doi [\mnras]
  {10.1093/mnras/stv027}, \href
  {https://ui.adsabs.harvard.edu/abs/2015MNRAS.448.1007B} {448, 1007}

\bibitem[\protect\citeauthoryear{{Behrendt}, {Burkert}  \&
  {Schartmann}}{{Behrendt} et~al.}{2016}]{Behrendt16}
{Behrendt} M.,  {Burkert} A.,   {Schartmann} M.,  2016, \mn@doi [\apjl]
  {10.3847/2041-8205/819/1/L2}, \href
  {https://ui.adsabs.harvard.edu/abs/2016ApJ...819L...2B} {819, L2}

\bibitem[\protect\citeauthoryear{{Behrendt}, {Schartmann}  \&
  {Burkert}}{{Behrendt} et~al.}{2019}]{Behrendt19}
{Behrendt} M.,  {Schartmann} M.,   {Burkert} A.,  2019, \mn@doi [\mnras]
  {10.1093/mnras/stz1717}, \href
  {https://ui.adsabs.harvard.edu/abs/2019MNRAS.488..306B} {488, 306}

\bibitem[\protect\citeauthoryear{{Behrens}, {Pallottini}, {Ferrara},
  {Gallerani}  \& {Vallini}}{{Behrens} et~al.}{2018}]{Behrens18}
{Behrens} C.,  {Pallottini} A.,  {Ferrara} A.,  {Gallerani} S.,   {Vallini} L.,
   2018, \mn@doi [\mnras] {10.1093/mnras/sty552}, \href
  {https://ui.adsabs.harvard.edu/abs/2018MNRAS.477..552B} {477, 552}

\bibitem[\protect\citeauthoryear{{Bertoldi} \& {McKee}}{{Bertoldi} \&
  {McKee}}{1992}]{Bertoldi92}
{Bertoldi} F.,  {McKee} C.~F.,  1992, \mn@doi [\apj] {10.1086/171638}, \href
  {https://ui.adsabs.harvard.edu/abs/1992ApJ...395..140B} {395, 140}

\bibitem[\protect\citeauthoryear{{B{\'e}thermin} et~al.,}{{B{\'e}thermin}
  et~al.}{2012}]{Bethermin12}
{B{\'e}thermin} M.,  et~al., 2012, \mn@doi [\apjl]
  {10.1088/2041-8205/757/2/L23}, \href
  {https://ui.adsabs.harvard.edu/abs/2012ApJ...757L..23B} {757, L23}

\bibitem[\protect\citeauthoryear{{B{\'e}thermin} et~al.,}{{B{\'e}thermin}
  et~al.}{2015}]{Bethermin15}
{B{\'e}thermin} M.,  et~al., 2015, \mn@doi [\aap]
  {10.1051/0004-6361/201425031}, \href
  {https://ui.adsabs.harvard.edu/abs/2015A&A...573A.113B} {573, A113}

\bibitem[\protect\citeauthoryear{{Binney} \& {Tremaine}}{{Binney} \&
  {Tremaine}}{2008}]{Binney}
{Binney} J.,  {Tremaine} S.,  2008, {Galactic Dynamics: Second Edition}

\bibitem[\protect\citeauthoryear{{Bournaud}}{{Bournaud}}{2016}]{Bournaud16}
{Bournaud} F.,  2016, in {Laurikainen} E.,  {Peletier} R.,   {Gadotti} D.,
  eds,  Astrophysics and Space Science Library Vol. 418, Galactic Bulges.
  p.~355 (\mn@eprint {arXiv} {1503.07660}),
  \mn@doi{10.1007/978-3-319-19378-6_13}

\bibitem[\protect\citeauthoryear{{Bournaud} \& {Elmegreen}}{{Bournaud} \&
  {Elmegreen}}{2009}]{BE09}
{Bournaud} F.,  {Elmegreen} B.~G.,  2009, \mn@doi [\apjl]
  {10.1088/0004-637X/694/2/L158}, \href
  {https://ui.adsabs.harvard.edu/abs/2009ApJ...694L.158B} {694, L158}

\bibitem[\protect\citeauthoryear{{Bournaud}, {Elmegreen}  \&
  {Elmegreen}}{{Bournaud} et~al.}{2007}]{BEE07}
{Bournaud} F.,  {Elmegreen} B.~G.,   {Elmegreen} D.~M.,  2007, \mn@doi [\apj]
  {10.1086/522077}, \href
  {https://ui.adsabs.harvard.edu/abs/2007ApJ...670..237B} {670, 237}

\bibitem[\protect\citeauthoryear{{Bournaud} et~al.,}{{Bournaud}
  et~al.}{2008}]{Bournaud08}
{Bournaud} F.,  et~al., 2008, \mn@doi [\aap] {10.1051/0004-6361:20079250},
  \href {https://ui.adsabs.harvard.edu/abs/2008A&A...486..741B} {486, 741}

\bibitem[\protect\citeauthoryear{{Bournaud}, {Elmegreen}, {Teyssier}, {Block}
  \& {Puerari}}{{Bournaud} et~al.}{2010}]{Bournaud10}
{Bournaud} F.,  {Elmegreen} B.~G.,  {Teyssier} R.,  {Block} D.~L.,   {Puerari}
  I.,  2010, \mn@doi [\mnras] {10.1111/j.1365-2966.2010.17370.x}, \href
  {https://ui.adsabs.harvard.edu/abs/2010MNRAS.409.1088B} {409, 1088}

\bibitem[\protect\citeauthoryear{{Bournaud} et~al.,}{{Bournaud}
  et~al.}{2014}]{Bournaud14}
{Bournaud} F.,  et~al., 2014, \mn@doi [\apj] {10.1088/0004-637X/780/1/57},
  \href {https://ui.adsabs.harvard.edu/\#abs/2014ApJ...780...57B} {780, 57}

\bibitem[\protect\citeauthoryear{{Bruzual} \& {Charlot}}{{Bruzual} \&
  {Charlot}}{2003}]{Bruzual03}
{Bruzual} G.,  {Charlot} S.,  2003, \mn@doi [\mnras]
  {10.1046/j.1365-8711.2003.06897.x}, \href
  {https://ui.adsabs.harvard.edu/abs/2003MNRAS.344.1000B} {344, 1000}

\bibitem[\protect\citeauthoryear{{Cava}, {Schaerer}, {Richard},
  {P{\'e}rez-Gonz{\'a}lez}, {Dessauges-Zavadsky}, {Mayer}  \&
  {Tamburello}}{{Cava} et~al.}{2018}]{Cava18}
{Cava} A.,  {Schaerer} D.,  {Richard} J.,  {P{\'e}rez-Gonz{\'a}lez} P.~G.,
  {Dessauges-Zavadsky} M.,  {Mayer} L.,   {Tamburello} V.,  2018, \mn@doi
  [Nature Astronomy] {10.1038/s41550-017-0295-x}, \href
  {https://ui.adsabs.harvard.edu/\#abs/2018NatAs...2...76C} {2, 76}

\bibitem[\protect\citeauthoryear{{Ceverino}, {Dekel}  \& {Bournaud}}{{Ceverino}
  et~al.}{2010}]{Ceverino10}
{Ceverino} D.,  {Dekel} A.,   {Bournaud} F.,  2010, \mn@doi [\mnras]
  {10.1111/j.1365-2966.2010.16433.x}, \href
  {https://ui.adsabs.harvard.edu/abs/2010MNRAS.404.2151C} {404, 2151}

\bibitem[\protect\citeauthoryear{{Ceverino}, {Dekel}, {Mandelker}, {Bournaud},
  {Burkert}, {Genzel}  \& {Primack}}{{Ceverino} et~al.}{2012}]{Ceverino12}
{Ceverino} D.,  {Dekel} A.,  {Mandelker} N.,  {Bournaud} F.,  {Burkert} A.,
  {Genzel} R.,   {Primack} J.,  2012, \mn@doi [\mnras]
  {10.1111/j.1365-2966.2011.20296.x}, \href
  {https://ui.adsabs.harvard.edu/abs/2012MNRAS.420.3490C} {420, 3490}

\bibitem[\protect\citeauthoryear{{Ceverino}, {Dekel}, {Tweed}  \&
  {Primack}}{{Ceverino} et~al.}{2015}]{Ceverino15}
{Ceverino} D.,  {Dekel} A.,  {Tweed} D.,   {Primack} J.,  2015, \mn@doi
  [\mnras] {10.1093/mnras/stu2694}, \href
  {http://adsabs.harvard.edu/abs/2015MNRAS.447.3291C} {447, 3291}

\bibitem[\protect\citeauthoryear{{Chabrier}}{{Chabrier}}{2003}]{Chabrier03}
{Chabrier} G.,  2003, \mn@doi [\pasp] {10.1086/376392}, \href
  {https://ui.adsabs.harvard.edu/abs/2003PASP..115..763C} {115, 763}

\bibitem[\protect\citeauthoryear{{Cibinel} et~al.,}{{Cibinel}
  et~al.}{2015}]{Cibinel15}
{Cibinel} A.,  et~al., 2015, \mn@doi [\apj] {10.1088/0004-637X/805/2/181},
  \href {https://ui.adsabs.harvard.edu/abs/2015ApJ...805..181C} {805, 181}

\bibitem[\protect\citeauthoryear{{Cibinel} et~al.,}{{Cibinel}
  et~al.}{2017}]{Cibinel17}
{Cibinel} A.,  et~al., 2017, \mn@doi [\mnras] {10.1093/mnras/stx1112}, \href
  {https://ui.adsabs.harvard.edu/abs/2017MNRAS.469.4683C} {469, 4683}

\bibitem[\protect\citeauthoryear{{Combes}, {Garc{\'\i}a-Burillo}, {Braine},
  {Schinnerer}, {Walter}  \& {Colina}}{{Combes} et~al.}{2013}]{Combes13}
{Combes} F.,  {Garc{\'\i}a-Burillo} S.,  {Braine} J.,  {Schinnerer} E.,
  {Walter} F.,   {Colina} L.,  2013, \mn@doi [\aap]
  {10.1051/0004-6361/201220392}, \href
  {https://ui.adsabs.harvard.edu/abs/2013A&A...550A..41C} {550, A41}

\bibitem[\protect\citeauthoryear{{Courty} \& {Alimi}}{{Courty} \&
  {Alimi}}{2004}]{Courty04}
{Courty} S.,  {Alimi} J.~M.,  2004, \mn@doi [\aap]
  {10.1051/0004-6361:20031736}, \href
  {https://ui.adsabs.harvard.edu/abs/2004A&A...416..875C} {416, 875}

\bibitem[\protect\citeauthoryear{{Cowie}, {Songaila}, {Hu}  \& {Cohen}}{{Cowie}
  et~al.}{1996}]{Cowie96}
{Cowie} L.~L.,  {Songaila} A.,  {Hu} E.~M.,   {Cohen} J.~G.,  1996, \mn@doi
  [\aj] {10.1086/118058}, \href
  {https://ui.adsabs.harvard.edu/abs/1996AJ....112..839C} {112, 839}

\bibitem[\protect\citeauthoryear{{Daddi} et~al.,}{{Daddi}
  et~al.}{2010}]{Daddi10}
{Daddi} E.,  et~al., 2010, \mn@doi [\apj] {10.1088/0004-637X/713/1/686}, \href
  {https://ui.adsabs.harvard.edu/abs/2010ApJ...713..686D} {713, 686}

\bibitem[\protect\citeauthoryear{{Dekel} \& {Krumholz}}{{Dekel} \&
  {Krumholz}}{2013}]{Dekel13}
{Dekel} A.,  {Krumholz} M.~R.,  2013, \mn@doi [\mnras] {10.1093/mnras/stt480},
  \href {http://adsabs.harvard.edu/abs/2013MNRAS.432..455D} {432, 455}

\bibitem[\protect\citeauthoryear{{Dekel}, {Sari}  \& {Ceverino}}{{Dekel}
  et~al.}{2009}]{DSC09}
{Dekel} A.,  {Sari} R.,   {Ceverino} D.,  2009, \mn@doi [\apj]
  {10.1088/0004-637X/703/1/785}, \href
  {https://ui.adsabs.harvard.edu/abs/2009ApJ...703..785D} {703, 785}

\bibitem[\protect\citeauthoryear{{Dessauges-Zavadsky}, {Schaerer}, {Cava},
  {Mayer}  \& {Tamburello}}{{Dessauges-Zavadsky} et~al.}{2017}]{Dessauges17}
{Dessauges-Zavadsky} M.,  {Schaerer} D.,  {Cava} A.,  {Mayer} L.,
  {Tamburello} V.,  2017, \mn@doi [\apj] {10.3847/2041-8213/aa5d52}, \href
  {https://ui.adsabs.harvard.edu/\#abs/2017ApJ...836L..22D} {836, L22}

\bibitem[\protect\citeauthoryear{{Dessauges-Zavadsky}
  et~al.,}{{Dessauges-Zavadsky} et~al.}{2019}]{Dessauges19}
{Dessauges-Zavadsky} M.,  et~al., 2019, \mn@doi [Nature Astronomy]
  {10.1038/s41550-019-0874-0}, \href
  {https://ui.adsabs.harvard.edu/abs/2019NatAs.tmp..436D} {p.~436}

\bibitem[\protect\citeauthoryear{{Dubois} \& {Teyssier}}{{Dubois} \&
  {Teyssier}}{2008}]{Dubois08}
{Dubois} Y.,  {Teyssier} R.,  2008, \mn@doi [\aap]
  {10.1051/0004-6361:20078326}, \href
  {https://ui.adsabs.harvard.edu/abs/2008A&A...477...79D} {477, 79}

\bibitem[\protect\citeauthoryear{{Dutton} et~al.,}{{Dutton}
  et~al.}{2011}]{Dutton11}
{Dutton} A.~A.,  et~al., 2011, \mn@doi [\mnras]
  {10.1111/j.1365-2966.2010.17555.x}, \href
  {https://ui.adsabs.harvard.edu/abs/2011MNRAS.410.1660D} {410, 1660}

\bibitem[\protect\citeauthoryear{{Ebeling}, {Ma}, {Kneib}, {Jullo}, {Courtney},
  {Barrett}, {Edge}  \& {Le Borgne}}{{Ebeling} et~al.}{2009}]{Ebeling09}
{Ebeling} H.,  {Ma} C.~J.,  {Kneib} J.~P.,  {Jullo} E.,  {Courtney} N.~J.~D.,
  {Barrett} E.,  {Edge} A.~C.,   {Le Borgne} J.~F.,  2009, \mn@doi [\mnras]
  {10.1111/j.1365-2966.2009.14502.x}, \href
  {https://ui.adsabs.harvard.edu/\#abs/2009MNRAS.395.1213E} {395, 1213}

\bibitem[\protect\citeauthoryear{{Elbaz} et~al.,}{{Elbaz}
  et~al.}{2011}]{Elbaz2011}
{Elbaz} D.,  et~al., 2011, \mn@doi [\aap] {10.1051/0004-6361/201117239}, \href
  {https://ui.adsabs.harvard.edu/abs/2011A&A...533A.119E} {533, A119}

\bibitem[\protect\citeauthoryear{{Elmegreen} \& {Elmegreen}}{{Elmegreen} \&
  {Elmegreen}}{2005}]{EE05}
{Elmegreen} B.~G.,  {Elmegreen} D.~M.,  2005, \mn@doi [\apj] {10.1086/430514},
  \href {https://ui.adsabs.harvard.edu/\#abs/2005ApJ...627..632E} {627, 632}

\bibitem[\protect\citeauthoryear{{Elmegreen}, {Elmegreen}, {Vollbach}, {Foster}
   \& {Ferguson}}{{Elmegreen} et~al.}{2005}]{Elmegreen05}
{Elmegreen} B.~G.,  {Elmegreen} D.~M.,  {Vollbach} D.~R.,  {Foster} E.~R.,
  {Ferguson} T.~E.,  2005, \mn@doi [\apj] {10.1086/496952}, \href
  {https://ui.adsabs.harvard.edu/abs/2005ApJ...634..101E} {634, 101}

\bibitem[\protect\citeauthoryear{{Elmegreen}, {Elmegreen}, {Ravindranath}  \&
  {Coe}}{{Elmegreen} et~al.}{2007}]{Elmegreen07}
{Elmegreen} D.~M.,  {Elmegreen} B.~G.,  {Ravindranath} S.,   {Coe} D.~A.,
  2007, \mn@doi [\apj] {10.1086/511667}, \href
  {https://ui.adsabs.harvard.edu/\#abs/2007ApJ...658..763E} {658, 763}

\bibitem[\protect\citeauthoryear{{Fensch} \& {Bournaud}}{{Fensch} \&
  {Bournaud}}{2020}]{Fensch20}
{Fensch} J.,  {Bournaud} F.,  2020, arXiv e-prints, \href
  {https://ui.adsabs.harvard.edu/abs/2020arXiv201112966F} {p. arXiv:2011.12966}

\bibitem[\protect\citeauthoryear{{Fisher} et~al.,}{{Fisher}
  et~al.}{2017}]{Fisher17}
{Fisher} D.~B.,  et~al., 2017, \mn@doi [\mnras] {10.1093/mnras/stw2281}, \href
  {https://ui.adsabs.harvard.edu/abs/2017MNRAS.464..491F} {464, 491}

\bibitem[\protect\citeauthoryear{{Genzel} et~al.,}{{Genzel}
  et~al.}{2006}]{Genzel06}
{Genzel} R.,  et~al., 2006, \mn@doi [\nat] {10.1038/nature05052}, \href
  {https://ui.adsabs.harvard.edu/abs/2006Natur.442..786G} {442, 786}

\bibitem[\protect\citeauthoryear{{Genzel} et~al.,}{{Genzel}
  et~al.}{2008}]{Genzel08}
{Genzel} R.,  et~al., 2008, \mn@doi [\apj] {10.1086/591840}, \href
  {https://ui.adsabs.harvard.edu/\#abs/2008ApJ...687...59G} {687, 59}

\bibitem[\protect\citeauthoryear{{Guo} et~al.,}{{Guo} et~al.}{2018}]{Guo18}
{Guo} Y.,  et~al., 2018, \mn@doi [\apj] {10.3847/1538-4357/aaa018}, \href
  {https://ui.adsabs.harvard.edu/abs/2018ApJ...853..108G} {853, 108}

\bibitem[\protect\citeauthoryear{{G{\"u}ver} \& {{\"O}zel}}{{G{\"u}ver} \&
  {{\"O}zel}}{2009}]{Guver09}
{G{\"u}ver} T.,  {{\"O}zel} F.,  2009, \mn@doi [\mnras]
  {10.1111/j.1365-2966.2009.15598.x}, \href
  {https://ui.adsabs.harvard.edu/abs/2009MNRAS.400.2050G} {400, 2050}

\bibitem[\protect\citeauthoryear{{Haardt} \& {Madau}}{{Haardt} \&
  {Madau}}{1996}]{Haardt96}
{Haardt} F.,  {Madau} P.,  1996, \mn@doi [\apj] {10.1086/177035}, \href
  {https://ui.adsabs.harvard.edu/abs/1996ApJ...461...20H} {461, 20}

\bibitem[\protect\citeauthoryear{{Heywood}, {Avison}  \& {Williams}}{{Heywood}
  et~al.}{2011}]{Heywood11}
{Heywood} I.,  {Avison} A.,   {Williams} C.~J.,  2011, arXiv e-prints, \href
  {https://ui.adsabs.harvard.edu/abs/2011arXiv1106.3516H} {p. arXiv:1106.3516}

\bibitem[\protect\citeauthoryear{{Hopkins}, {Kere{\v s}}, {Murray}, {Quataert}
  \& {Hernquist}}{{Hopkins} et~al.}{2012}]{Hopkins12}
{Hopkins} P.~F.,  {Kere{\v s}} D.,  {Murray} N.,  {Quataert} E.,   {Hernquist}
  L.,  2012, \mn@doi [\mnras] {10.1111/j.1365-2966.2012.21981.x}, \href
  {http://adsabs.harvard.edu/abs/2012MNRAS.427..968H} {427, 968}

\bibitem[\protect\citeauthoryear{{Ivison}, {Richard}, {Biggs}, {Zwaan},
  {Falgarone}, {Arumugam}, {van der Werf}  \& {Rujopakarn}}{{Ivison}
  et~al.}{2020}]{Ivison20}
{Ivison} R.~J.,  {Richard} J.,  {Biggs} A.~D.,  {Zwaan} M.~A.,  {Falgarone} E.,
   {Arumugam} V.,  {van der Werf} P.~P.,   {Rujopakarn} W.,  2020, \mn@doi
  [\mnras] {10.1093/mnrasl/slaa046}, \href
  {https://ui.adsabs.harvard.edu/abs/2020MNRAS.495L...1I} {495, L1}

\bibitem[\protect\citeauthoryear{{Jeans}}{{Jeans}}{1902}]{Jeans1902}
{Jeans} J.~H.,  1902, \mn@doi [Philosophical Transactions of the Royal Society
  of London Series A] {10.1098/rsta.1902.0012}, \href
  {https://ui.adsabs.harvard.edu/abs/1902RSPTA.199....1J} {199, 1}

\bibitem[\protect\citeauthoryear{{Jullo}, {Kneib}, {Limousin},
  {El{\'\i}asd{\'o}ttir}, {Marshall}  \& {Verdugo}}{{Jullo}
  et~al.}{2007}]{Jullo07}
{Jullo} E.,  {Kneib} J.~P.,  {Limousin} M.,  {El{\'\i}asd{\'o}ttir} {\'A}.,
  {Marshall} P.~J.,   {Verdugo} T.,  2007, \mn@doi [New Journal of Physics]
  {10.1088/1367-2630/9/12/447}, \href
  {https://ui.adsabs.harvard.edu/abs/2007NJPh....9..447J} {9, 447}

\bibitem[\protect\citeauthoryear{{Kim}, {Ostriker}  \& {Stone}}{{Kim}
  et~al.}{2002}]{Kim02}
{Kim} W.-T.,  {Ostriker} E.~C.,   {Stone} J.~M.,  2002, \mn@doi [\apj]
  {10.1086/344367}, \href
  {https://ui.adsabs.harvard.edu/abs/2002ApJ...581.1080K} {581, 1080}

\bibitem[\protect\citeauthoryear{{Krumholz} \& {Dekel}}{{Krumholz} \&
  {Dekel}}{2010}]{Krumholz10}
{Krumholz} M.~R.,  {Dekel} A.,  2010, \mn@doi [\mnras]
  {10.1111/j.1365-2966.2010.16675.x}, \href
  {https://ui.adsabs.harvard.edu/abs/2010MNRAS.406..112K} {406, 112}

\bibitem[\protect\citeauthoryear{{Leung}, {Pallottini}, {Ferrara}  \& {Mac
  Low}}{{Leung} et~al.}{2020}]{Leung20}
{Leung} T.~K.~D.,  {Pallottini} A.,  {Ferrara} A.,   {Mac Low} M.-M.,  2020,
  \mn@doi [\apj] {10.3847/1538-4357/ab8cbb}, \href
  {https://ui.adsabs.harvard.edu/abs/2020ApJ...895...24L} {895, 24}

\bibitem[\protect\citeauthoryear{{Magdis} et~al.,}{{Magdis}
  et~al.}{2012}]{Magdis12}
{Magdis} G.~E.,  et~al., 2012, \mn@doi [\apj] {10.1088/0004-637X/760/1/6},
  \href {https://ui.adsabs.harvard.edu/abs/2012ApJ...760....6M} {760, 6}

\bibitem[\protect\citeauthoryear{{Murray}, {Quataert}  \& {Thompson}}{{Murray}
  et~al.}{2010}]{Murray10}
{Murray} N.,  {Quataert} E.,   {Thompson} T.~A.,  2010, \mn@doi [\apj]
  {10.1088/0004-637X/709/1/191}, \href
  {https://ui.adsabs.harvard.edu/abs/2010ApJ...709..191M} {709, 191}

\bibitem[\protect\citeauthoryear{{Noguchi}}{{Noguchi}}{1999}]{Noguchi99}
{Noguchi} M.,  1999, \mn@doi [\apj] {10.1086/306932}, \href
  {https://ui.adsabs.harvard.edu/abs/1999ApJ...514...77N} {514, 77}

\bibitem[\protect\citeauthoryear{{Oklop{\v{c}}i{\'c}}, {Hopkins}, {Feldmann},
  {Kere{\v{s}}}, {Faucher-Gigu{\`e}re}  \& {Murray}}{{Oklop{\v{c}}i{\'c}}
  et~al.}{2017}]{Oklopcic17}
{Oklop{\v{c}}i{\'c}} A.,  {Hopkins} P.~F.,  {Feldmann} R.,  {Kere{\v{s}}} D.,
  {Faucher-Gigu{\`e}re} C.-A.,   {Murray} N.,  2017, \mn@doi [\mnras]
  {10.1093/mnras/stw2754}, \href
  {https://ui.adsabs.harvard.edu/\#abs/2017MNRAS.465..952O} {465, 952}

\bibitem[\protect\citeauthoryear{{Parmentier} \& {Gilmore}}{{Parmentier} \&
  {Gilmore}}{2005}]{Parmentier05}
{Parmentier} G.,  {Gilmore} G.,  2005, \mn@doi [\mnras]
  {10.1111/j.1365-2966.2005.09455.x}, \href
  {https://ui.adsabs.harvard.edu/abs/2005MNRAS.363..326P} {363, 326}

\bibitem[\protect\citeauthoryear{{Perret}, {Renaud}, {Epinat}, {Amram},
  {Bournaud}, {Contini}, {Teyssier}  \& {Lambert}}{{Perret}
  et~al.}{2014}]{Perret14}
{Perret} V.,  {Renaud} F.,  {Epinat} B.,  {Amram} P.,  {Bournaud} F.,
  {Contini} T.,  {Teyssier} R.,   {Lambert} J.~C.,  2014, \mn@doi [\aap]
  {10.1051/0004-6361/201322395}, \href
  {https://ui.adsabs.harvard.edu/abs/2014A&A...562A...1P} {562, A1}

\bibitem[\protect\citeauthoryear{{Renaud} et~al.,}{{Renaud}
  et~al.}{2013}]{Renaud13}
{Renaud} F.,  et~al., 2013, \mn@doi [\mnras] {10.1093/mnras/stt1698}, \href
  {https://ui.adsabs.harvard.edu/abs/2013MNRAS.436.1836R} {436, 1836}

\bibitem[\protect\citeauthoryear{{Renaud}, {Bournaud}  \& {Duc}}{{Renaud}
  et~al.}{2015}]{Renaud15}
{Renaud} F.,  {Bournaud} F.,   {Duc} P.-A.,  2015, \mn@doi [\mnras]
  {10.1093/mnras/stu2208}, \href
  {https://ui.adsabs.harvard.edu/abs/2015MNRAS.446.2038R} {446, 2038}

\bibitem[\protect\citeauthoryear{{Rujopakarn} et~al.,}{{Rujopakarn}
  et~al.}{2019}]{Rujopakarn19}
{Rujopakarn} W.,  et~al., 2019, arXiv e-prints, \href
  {https://ui.adsabs.harvard.edu/abs/2019arXiv190404507R} {p. arXiv:1904.04507}

\bibitem[\protect\citeauthoryear{{Salpeter}}{{Salpeter}}{1955}]{Salpeter55}
{Salpeter} E.~E.,  1955, \mn@doi [\apj] {10.1086/145971}, \href
  {https://ui.adsabs.harvard.edu/abs/1955ApJ...121..161S} {121, 161}

\bibitem[\protect\citeauthoryear{{Santini} et~al.,}{{Santini}
  et~al.}{2014}]{Santini14}
{Santini} P.,  et~al., 2014, \mn@doi [\aap] {10.1051/0004-6361/201322835},
  \href {https://ui.adsabs.harvard.edu/abs/2014A&A...562A..30S} {562, A30}

\bibitem[\protect\citeauthoryear{{Schreiber} et~al.,}{{Schreiber}
  et~al.}{2015}]{Schreiber15}
{Schreiber} C.,  et~al., 2015, \mn@doi [\aap] {10.1051/0004-6361/201425017},
  \href {https://ui.adsabs.harvard.edu/abs/2015A&A...575A..74S} {575, A74}

\bibitem[\protect\citeauthoryear{{Shapiro}, {Genzel}  \& {F{\"o}rster
  Schreiber}}{{Shapiro} et~al.}{2010}]{Shapiro10}
{Shapiro} K.~L.,  {Genzel} R.,   {F{\"o}rster Schreiber} N.~M.,  2010, \mn@doi
  [\mnras] {10.1111/j.1745-3933.2010.00810.x}, \href
  {https://ui.adsabs.harvard.edu/abs/2010MNRAS.403L..36S} {403, L36}

\bibitem[\protect\citeauthoryear{{Swinbank} et~al.,}{{Swinbank}
  et~al.}{2009}]{Swinbank09}
{Swinbank} A.~M.,  et~al., 2009, \mn@doi [\mnras]
  {10.1111/j.1365-2966.2009.15617.x}, \href
  {https://ui.adsabs.harvard.edu/abs/2009MNRAS.400.1121S} {400, 1121}

\bibitem[\protect\citeauthoryear{{Swinbank} et~al.,}{{Swinbank}
  et~al.}{2010}]{Swinbank10}
{Swinbank} A.~M.,  et~al., 2010, \mn@doi [\nat] {10.1038/nature08880}, \href
  {https://ui.adsabs.harvard.edu/abs/2010Natur.464..733S} {464, 733}

\bibitem[\protect\citeauthoryear{{Tacconi} et~al.,}{{Tacconi}
  et~al.}{2010}]{Tacconi10}
{Tacconi} L.~J.,  et~al., 2010, \mn@doi [\nat] {10.1038/nature08773}, \href
  {https://ui.adsabs.harvard.edu/abs/2010Natur.463..781T} {463, 781}

\bibitem[\protect\citeauthoryear{{Tamburello}, {Mayer}, {Shen}  \&
  {Wadsley}}{{Tamburello} et~al.}{2015}]{Tamburello15}
{Tamburello} V.,  {Mayer} L.,  {Shen} S.,   {Wadsley} J.,  2015, \mn@doi
  [\mnras] {10.1093/mnras/stv1695}, \href
  {https://ui.adsabs.harvard.edu/abs/2015MNRAS.453.2490T} {453, 2490}

\bibitem[\protect\citeauthoryear{{Teyssier}}{{Teyssier}}{2002}]{Teyssier02}
{Teyssier} R.,  2002, \mn@doi [\aap] {10.1051/0004-6361:20011817}, \href
  {http://adsabs.harvard.edu/abs/2002A%26A...385..337T} {385, 337}

\bibitem[\protect\citeauthoryear{{Teyssier}, {Chapon}  \&
  {Bournaud}}{{Teyssier} et~al.}{2010}]{Teyssier10}
{Teyssier} R.,  {Chapon} D.,   {Bournaud} F.,  2010, \mn@doi [\apjl]
  {10.1088/2041-8205/720/2/L149}, \href
  {https://ui.adsabs.harvard.edu/abs/2010ApJ...720L.149T} {720, L149}

\bibitem[\protect\citeauthoryear{{Toomre}}{{Toomre}}{1964}]{Toomre64}
{Toomre} A.,  1964, \mn@doi [\apj] {10.1086/147861}, \href
  {https://ui.adsabs.harvard.edu/abs/1964ApJ...139.1217T} {139, 1217}

\bibitem[\protect\citeauthoryear{{Truelove}, {Klein}, {McKee}, {Holliman},
  {Howell}  \& {Greenough}}{{Truelove} et~al.}{1997}]{Truelove97}
{Truelove} J.~K.,  {Klein} R.~I.,  {McKee} C.~F.,  {Holliman} John~H. I.,
  {Howell} L.~H.,   {Greenough} J.~A.,  1997, \mn@doi [\apj] {10.1086/310975},
  \href {https://ui.adsabs.harvard.edu/abs/1997ApJ...489L.179T} {489, L179}

\bibitem[\protect\citeauthoryear{{Zanella} et~al.,}{{Zanella}
  et~al.}{2018}]{Zanella18}
{Zanella} A.,  et~al., 2018, \mn@doi [\mnras] {10.1093/mnras/sty2394}, \href
  {https://ui.adsabs.harvard.edu/abs/2018MNRAS.481.1976Z} {481, 1976}

\bibitem[\protect\citeauthoryear{{Zanella} et~al.,}{{Zanella}
  et~al.}{2019}]{Zanella19}
{Zanella} A.,  et~al., 2019, \mn@doi [\mnras] {10.1093/mnras/stz2099}, \href
  {https://ui.adsabs.harvard.edu/abs/2019MNRAS.489.2792Z} {489, 2792}

\bibitem[\protect\citeauthoryear{{Zanella}, {Pallottini}, {Ferrara},
  {Gallerani}, {Carniani}, {Kohandel}  \& {Behrens}}{{Zanella}
  et~al.}{2021}]{Zanella21}
{Zanella} A.,  {Pallottini} A.,  {Ferrara} A.,  {Gallerani} S.,  {Carniani} S.,
   {Kohandel} M.,   {Behrens} C.,  2021, \mn@doi [\mnras]
  {10.1093/mnras/staa2776}, \href
  {https://ui.adsabs.harvard.edu/abs/2021MNRAS.500..118Z} {500, 118}

\bibitem[\protect\citeauthoryear{{de Blok}, {Walter}, {Brinks}, {Trachternach},
  {Oh}  \& {Kennicutt}}{{de Blok} et~al.}{2008}]{deBlok08}
{de Blok} W.~J.~G.,  {Walter} F.,  {Brinks} E.,  {Trachternach} C.,  {Oh}
  S.~H.,   {Kennicutt} R.~C. J.,  2008, \mn@doi [\aj]
  {10.1088/0004-6256/136/6/2648}, \href
  {https://ui.adsabs.harvard.edu/abs/2008AJ....136.2648D} {136, 2648}

\makeatother
\end{thebibliography}

\newpage
\onecolumn

\appendix

\section{Detailed clumps mass per simulation}

\begin{figure*}
	\begin{center}
   		\includegraphics[width=\textwidth]{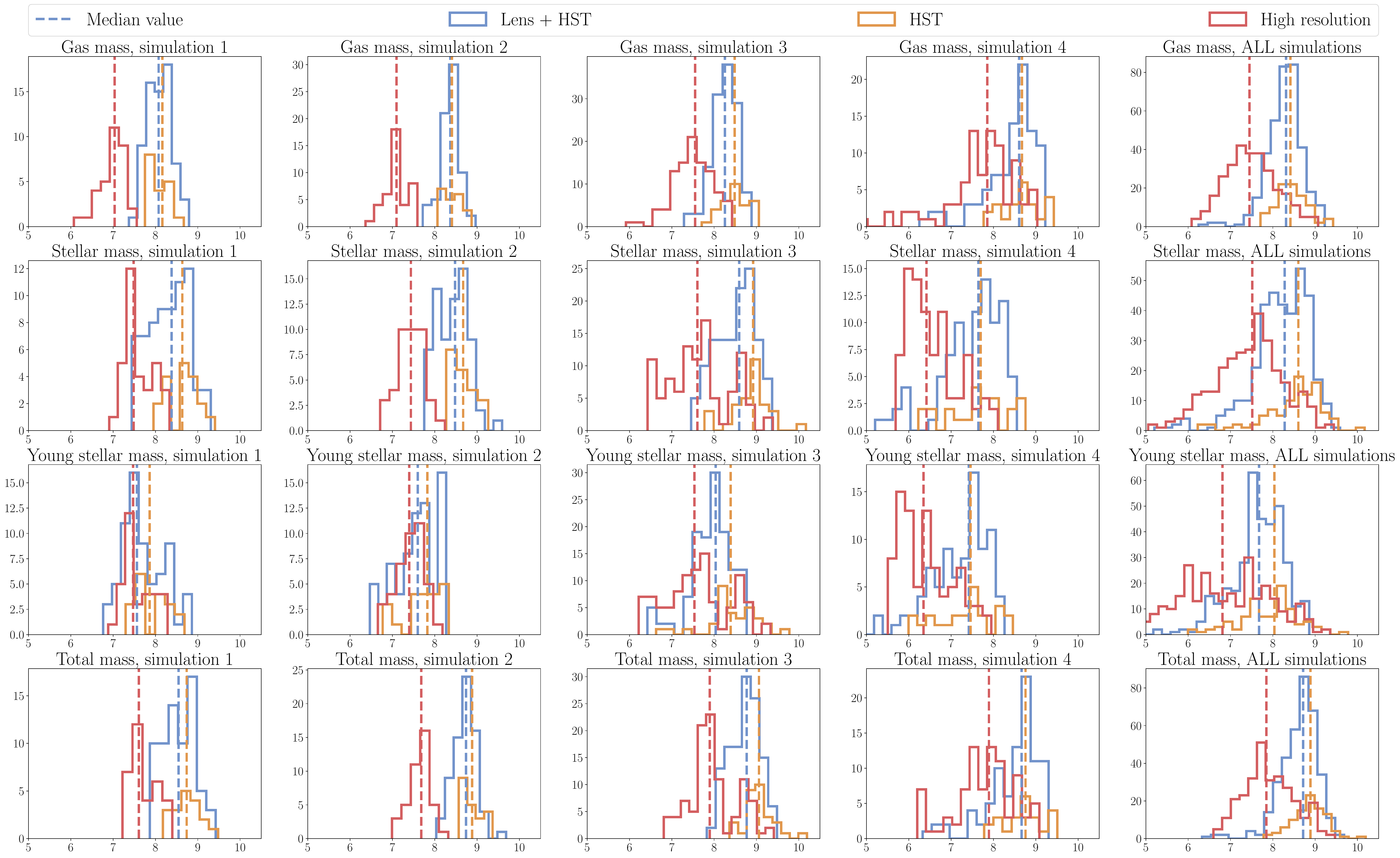}
		\caption{The top row represents the gas mass of the clumps found in the HST mock observations (orange histogram) in the lensed HST mock observations (blue histogram) and in High Resolution mock observations (red histogram). The second row is the stellar mass of the clumps. The third one is the mass of stars younger than 10 Myr. The fourth is the total mass (gas + stars). To each column corresponds a typical output of each simulation and the fifth one is the stacking of all four simulations. In each panel the dashed line is the median value of the corresponding histogram.}
		\label{bigplot}
	\end{center}
\end{figure*}

\newpage

\section{Chabrier and Salpeter IMF comparison.}

\begin{figure*}
    \centering
    \includegraphics[width=\textwidth]{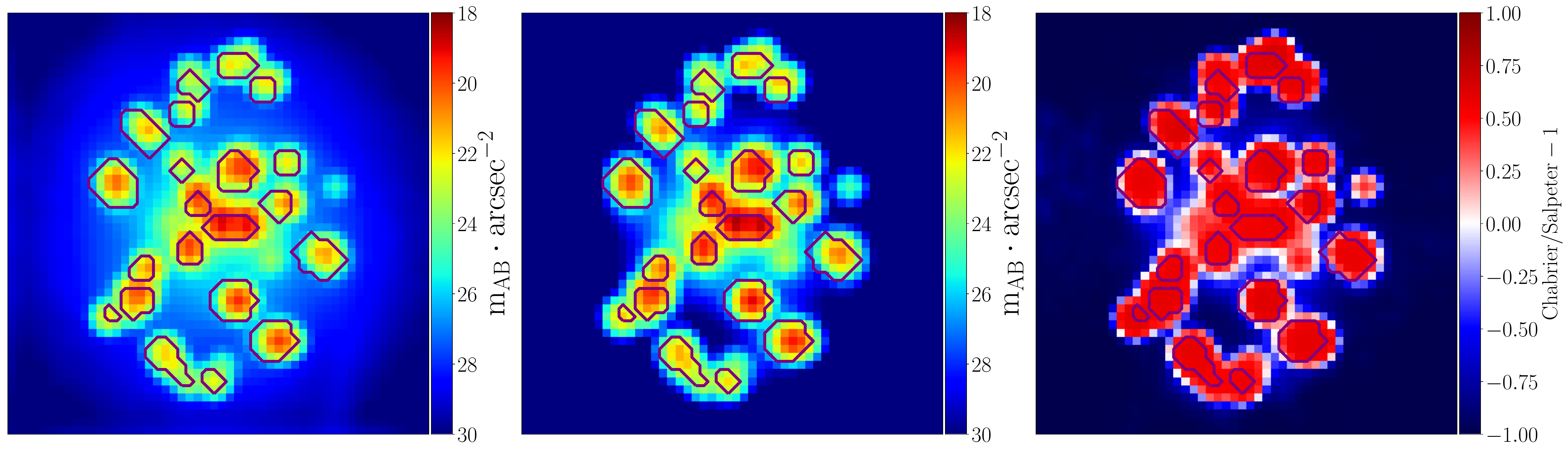}
    \caption{Comparison of Salpeter IMF (left) with Chabrier IMF (center) at HST resolution. A direct comparison is made on the right plot. The purple contour are the clumps detected by Astrodendro on the Salpeter IMF mock.}
    \label{fig:lr_chab}
\end{figure*}

\begin{figure*}
    \centering
    \includegraphics[width=\textwidth]{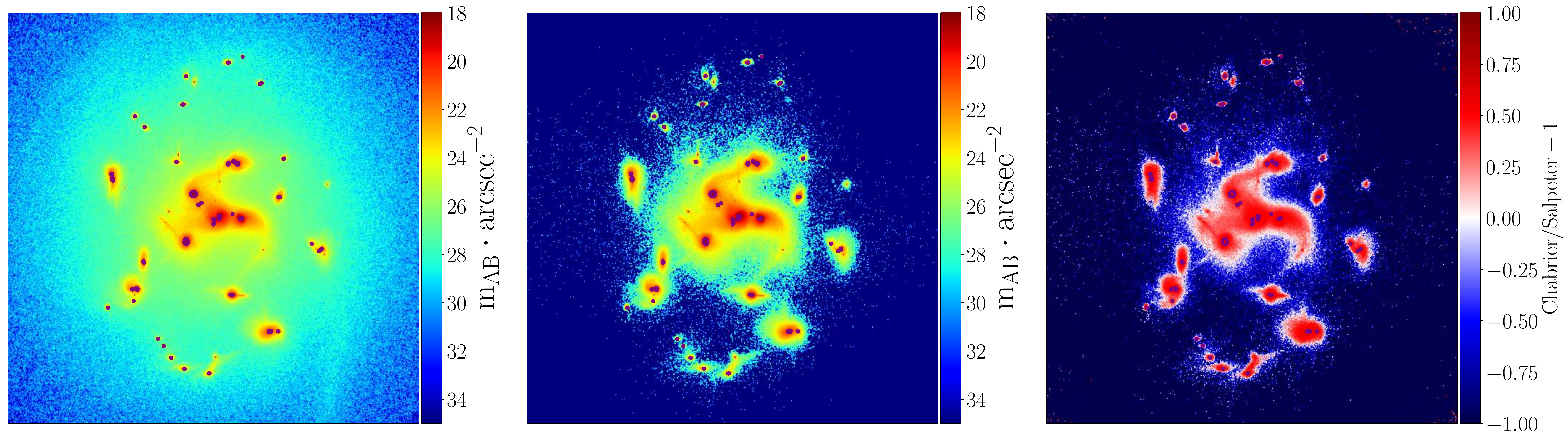}
    \caption{Comparison of Salpeter IMF (left) with Chabrier IMF (center) at high resolution. A direct comparison is made on the right plot. The purple contour are the clumps detected by Astrodendro on the Salpeter IMF mock.}
    \label{fig:hr_chab}
\end{figure*}

\begin{figure*}
    \centering
    \includegraphics[width=\textwidth]{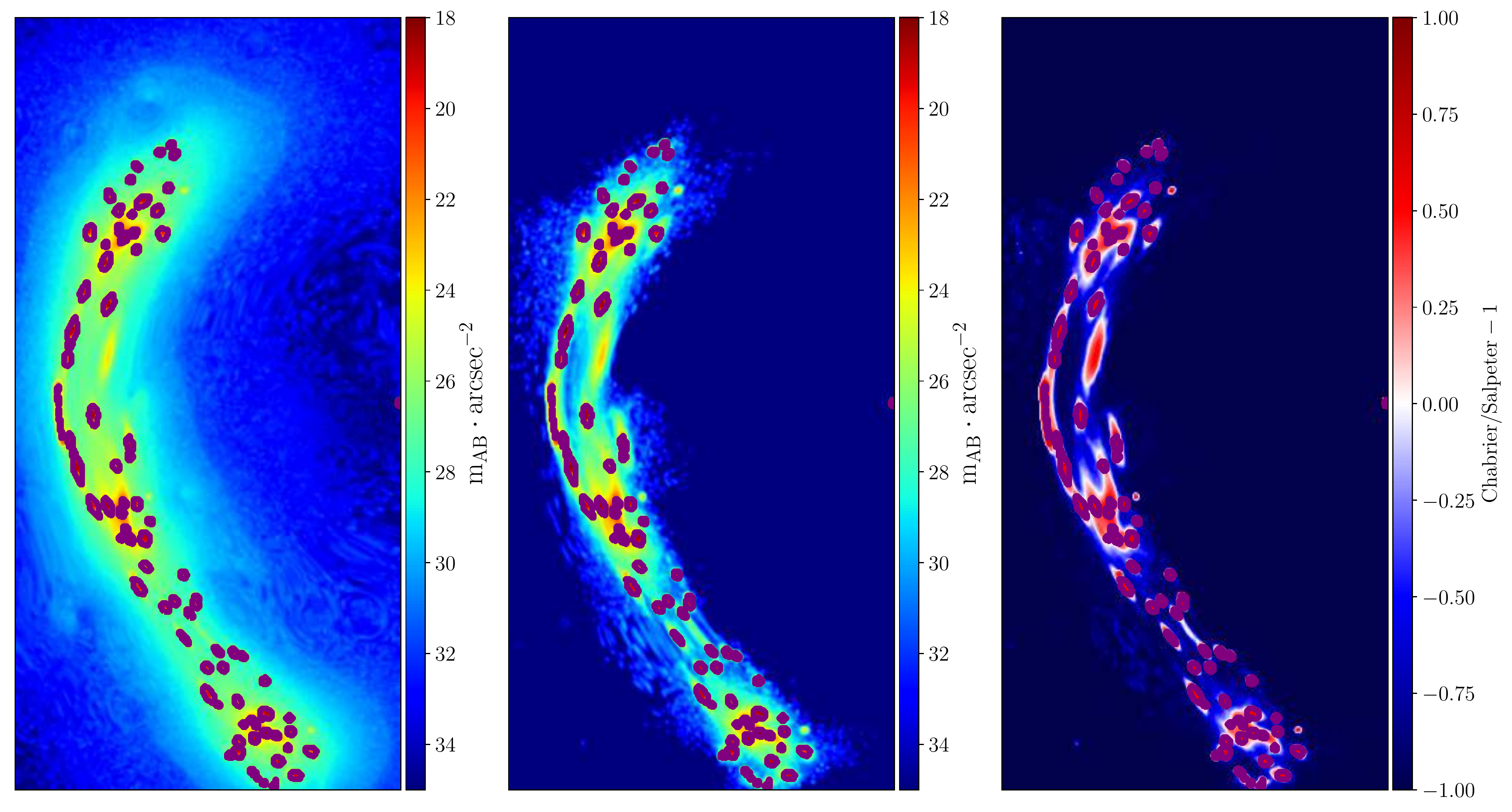}
    \caption{Comparison of Salpeter IMF (left) with Chabrier IMF (center) on lensed mocks. A direct comparison is made on the right plot. The purple contour are the clumps detected by Astrodendro on the Salpeter IMF mock.}
    \label{fig:lens_chab}
\end{figure*}

This appendix presents a brief comparison of different IMF for our simulations. Indeed our simulations do not resolve the stars in detail so one need to assume an IMF to know the mass distribution of the stars. A \cite{Chabrier03} IMF will have less low mass stars and more high mass than a \cite{Salpeter55} thus leading to a stronger contrast between low and high stellar density regions, so more contrasted giant clumps. Nevertheless by computing the mocks at the HST resolution and plotting the clumps detected on the mocks with a \cite{Salpeter55} IMF, as in Figure \ref{fig:lr_chab}, one can see that they are visually matching the giant clumps seen on mocks with a \cite{Chabrier03} IMF. This leads us to claim that in the scenario, the giant clumps detection is dominated by the clump finder parameters and not by the choice of the IMF.
The exact same effect can be seen with the High Resolution mocks, Figure \ref{fig:hr_chab}, and with the lens mocks, Figure \ref{fig:lens_chab}.
\label{app:IMF}

\end{document}